**Title: Modelling continual learning in humans with Hebbian context gating and exponentially decaying task signals**

**Short title: A neural network model of human continual learning**

Timo Flesch[1, ¶], David G. Nagy[2, ¶], Andrew Saxe[3,4], Christopher Summerfield[1]

[1] Department of Experimental Psychology, University of Oxford; Oxford OX2 6GG, UK.
[2] Department of Computational Sciences, Wigner Research Centre for Physics; Budapest, Hungary.
[3] Gatsby Computational Neuroscience Unit & Sainsbury Wellcome Centre, University College London; London, UK.
[4] CIFAR Azrieli Global Scholars program, CIFAR; Toronto, Canada.

*Corresponding author

E-mail: timo.flesch@gmail.com (TF)

¶These authors contributed equally to this work

**Abstract**

Humans can learn several tasks in succession with minimal mutual interference but perform more poorly when trained on multiple tasks at once. The opposite is true for standard deep neural networks. Here, we propose novel computational constraints for artificial neural networks, inspired by earlier work on gating in the primate prefrontal cortex, that capture the cost of interleaved training and allow the network to learn two tasks in sequence without forgetting. We augment standard stochastic gradient descent with two algorithmic motifs, so-called "sluggish" task units and a Hebbian training step that strengthens connections between task units and hidden units that encode task-relevant information. We found that the "sluggish" units introduce a switch-cost during training, which biases representations under interleaved training towards a joint representation that ignores the contextual cue, while the Hebbian step promotes the formation of a gating scheme from task units to the hidden layer that produces orthogonal representations which are perfectly guarded against interference. Validating the model on previously published human behavioural data revealed that it matches performance of participants who had been trained on blocked or interleaved curricula, and that these performance differences were driven by misestimation of the true category boundary.

**Author Summary**

Humans can learn multiple tasks over their lifetime with minimal forgetting. In contrast, machine learning architectures based on artificial neural networks fail to learn multiple tasks in sequence and require data of all tasks to be present at once. Previous reports suggest that the opposite is true for humans: We learn better when trained on one task at a time. Here, we sought to identify the basis-set of algorithmic motifs required to mimic human-like continual learning. We propose a novel training method inspired by insights into the function of prefrontal cortex of the human brain. The method consists of task neurons that carry information over successive trials, and an update step that links those to other neurons that encode task-relevant information. Together, these two innovations allow us to model human continual task performance. Analysing how the network represented task information revealed striking similarities between our network and recent reports on task representations in the prefrontal cortex of the mammalian brain. Taken together, our approach describes an effort to bridge insights from machine learning and neuroscience to advance our understanding of the algorithmic basis of continual learning.

# INTRODUCTION

Humans have the remarkable ability to learn multiple tasks over their lifespan. New tasks can be learned in sequence with minimal disruption to previously acquired tasks, a feat that is known as *continual learning*. For example, in the case of (supervised) visual categorisation, if so asked you could learn to successfully categorise fruits by size (crab apple vs. granny smith) and then by colour (ripe vs. unripe) without the latter learning overwriting the former. Building neural networks that learn continually has proved challenging in AI research [1,2]. In neuroscience, it remains an open question how the human brain learns continually, and whether biology can inspire candidate solutions for artificial agents [3–7]. Here, we present a computational model of human continual learning, which builds on earlier work on cognitive control and the neural basis of representation learning.

With the advent of deep learning, artificial neural networks are enjoying a renaissance as models of biological information processing [8,9]. Despite their architectural simplicity, representations that emerge in neural networks bear striking similarities to those observed in early visual cortex and higher association areas of the human brain [10–14], leading to the proposal that these models can be used as a test-bed for theories about the geometry [15,16] and dimensionality of neural representations [17,18], or the feature selectivity of downstream cortical areas [19]. However, without significant modification, neural networks trained with vanilla gradient descent fail at tests of continual learning: they are unable to learn multiple tasks in sequence without suffering from catastrophic forms of forgetting [1,20,21]. Interestingly, catastrophic interference is not restricted to simple feed-forward architectures trained on supervised learning problems but has also been observed in recurrent neural networks [22] and in the domain of reinforcement learning [23,24]. The reason for catastrophic forgetting is well understood, as standard deep learning approaches require training data to be independent and identically distributed (i.i.d.), a requirement that is violated when several tasks are learned in succession [20]. Crucially, however, some evidence suggests that humans may sometimes even perform worse when trained on i.i.d data [6,25], which challenges the assumption that standard deep learning architectures can serve as models of the learning dynamics observed in biological organisms.

Due to the ubiquitous nature of this problem, it has received considerable attention in the machine learning community, and numerous engineering solutions have been proposed that wholly or partially prevent forgetting [1,26,27], either by preventing task-relevant weights from changing [24,28], dynamic architecture growth [29], experience replay [23,30] or orthogonalization of representations in the hidden layer [31–33]. While initially devised as solutions for feed-forward models, many of these approaches have been successfully applied to recurrent neural networks [22,34,35] or other ML domains such as reinforcement learning [24].

Some solutions draw loose inspiration from neuroscience, such as *experience replay in reinforcement learning*, which can be related to Complementary Learning Systems theory [36,37] or *gating* approaches, linked to top-down attentional control [38], or *regularisation* approaches, which can be related to changes in synaptic plasticity on different timescales [39,40]. Orthogonalisation approaches in particular have gained attention in neuroscience [32,34], as more recent investigations of neural geometry have shown that during multi-task performance, mutual interference among tasks is minimised by projecting relevant dimensions into orthogonal, low-dimensional subspaces [15,17,41,42].

Here, rather than focussing on developing a novel solution, we used a computation modelling approach to understand how biological agents may learn multiple tasks in succession. We draw inspiration from research that has focussed on the implementation of control process in the prefrontal cortex. In neuroscience, the term "cognitive control" is applied to neural mechanisms that allow a context-appropriate task to be selected and executed with minimal interference [43]. Cognitive control has long been associated with the prefrontal cortex (PFC), based on evidence that prefrontal neurons code for specific tasks, and exert top-down control to prioritise context-appropriate stimuli and actions [43–48]. In the domain of categorisation, it has been proposed that the PFC may implement cognitive control by gating (or compressing) task-irrelevant input dimensions [49–51]. However, in classic models (such as that proposed by Miller and Cohen [43]) this gating process is implemented by hand. Here, instead, we asked how this gating process might be learned in a way that allows tasks to be learned with minimal mutual interference. We also draw upon other work that has proposed gating as a potential solution to continual learning, in both feedforward and recurrent neural networks, either as additive bias to the input of a hidden unit, or as multiplicative gate that acts on the unit's output [38,52–54]. Again however, in these papers, the gating signals were usually hard-coded. A key challenge, thus, is to identify a mechanism that can acquire this control signal in a continual learning setting and could account for the apparent costs associated with learning multiple tasks from i.i.d. data.

We sought to develop a neural network model inspired by theories on cognitive control that describes how humans learn to perform multiple categorisation tasks in series. A starting point for our work is the observation that humans actively benefit when categorisation tasks are temporally autocorrelated (blocked) during training. For example, consider a validation task which requires naturalistic stimuli (tree images) to be categorised alternately by dimensions of leaf and branch density. Humans benefit from a training regime consisting of long training blocks of unidimensional leafy or branchy rules, rather than training blocks in which leafy and branchy rules are interleaved together [25]. This benefit appears to be particularly pronounced when exemplars are highly heterogenous within and across tasks [6,7]. Thus, our goal was to identify a model that could learn from scratch to capture the benefit of blocking and the cost of interleaving, as well as the patterns of neural geometry that have been observed during multitask performance.

There are two key ideas that motivate our model design. The first is that biological neural circuits have intrinsic time constants of integration which ensure that decisions are driven by information from the immediate past as well as the present. This principle underlies ubiquitously observed trial history effects in decision tasks [55–57]. The second is that simple learning based on coincidence detection (such as Hebbian learning) allow groupings of inputs to be effectively orthogonalised. Our model capitalises on these principles by combining two algorithmic motifs. Firstly, we assume that neuronal responses are "sluggish": on each trial, inputs to the network contain some information carried over from previous trials. Carrying over contextual cues from previous trials increases task interference (switch costs) in interleaved conditions (where sequential trials may require performance of conflicting tasks) but not in blocked conditions (where sequential trials involve the same task). Secondly, we propose that a Hebbian learning step follows each supervised parameter update, to strengthen connections between task signalling units and hidden units that encode task-relevant information. This has the effect of orthogonalising the weights linking context to hidden units for the two tasks, allowing tasks to be represented in independent subspaces in the hidden layer [58]. This intervention thus implements a form of context-dependent gating [49,50]. However, in contrast to earlier work on cognitive control and related papers that have used gating as a means for continual learning [38,51,54], we demonstrate that this control signal can be acquired by a simple biologically-inspired mechanism and without direct intervention by the experimenter. Finally, we show that this model forms highly task-specific neural codes, similar to those reported in a series of recent studies on the geometry of representations in human and macaque prefrontal cortex [15,17,41,42].

## RESULTS

All simulations described here were developed to model relative performance on a context-dependent categorisation task following blocked and interleaved training. These results have been reported in an earlier manuscript, where we subjected human participants to a variant of the well-established context-dependent decision making task [25]. Here, we reanalysed this behavioural dataset. In the original task, participants were asked to decide what type of tree would grow well in two different gardens, which we called the north and south garden respectively (**Fig. 1A**). Unbeknownst to them a priori, trees varied parametrically in terms of their density of branches ("branchiness") and leaves ("leafiness") and only one of the two dimensions was relevant in each task and determined "growth success", indicated by a numerical reward/penalty that was associated with planting the tree in a specific garden (**Fig. 1B**). Each training trial began with an image of either the north or south garden, which served as contextual cue. This was followed by an image of a tree, which participants could choose to plant ("accept") or not to plant ("reject") (**Fig. 1C**). On training trials, participants would then receive a numerical reward, that depended on the level of leafiness in the north garden and the level of branchiness in the south garden. Participants were either trained continually on a "blocked" curriculum, or in an "interleaved" curriculum, where trials from both contexts were randomly interspersed. Both groups were evaluated on an interleaved test block without feedback (**Fig. 1D**).

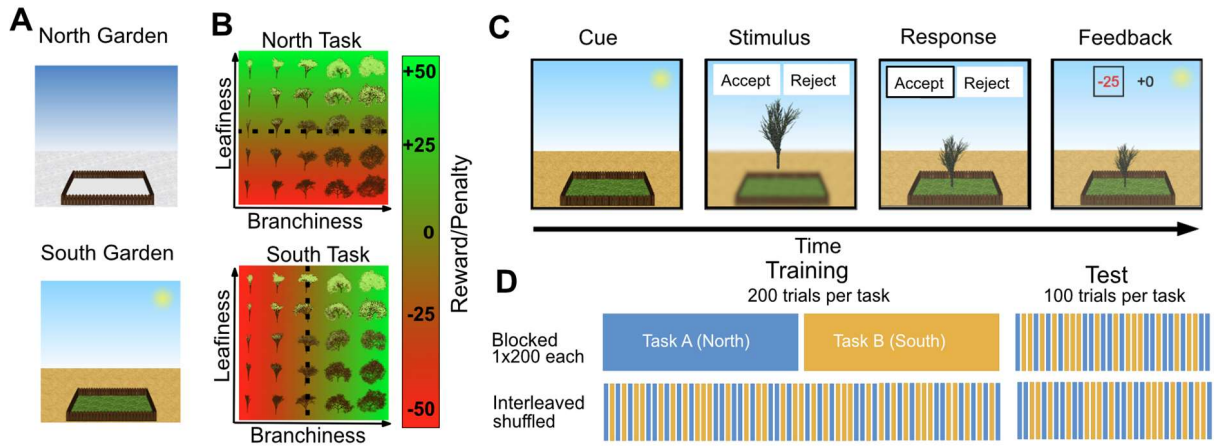

**Figure 1. Task design used in** [25]. (**A**) Contextual cues. The two contexts were illustrated as images of gardens, located either in a snowy (north garden) or desert-like environment (south garden). Participants were asked to learn which type of trees would grow well (i.e. give a reward for accepting them) in each of the two gardens. (**B**) Stimulus space and rules. Stimuli were procedurally generated fractal images of trees that varied parametrically in their density of leaves (leafiness) and branches (branchiness), spanning a 5x5 grid of possible feature combinations. Participants were asked to learn a context-dependent mapping from those trees to rewards associated with either accepting or rejecting them on a trial-by-trial basis. In each of two tasks (called the "north" and "south" tasks), only one of the two feature dimensions was relevant and determined the reward/penalty received for "accepting" a tree. (**C**) Trial structure. Each trial began with the display of a contextual cue that remained on the screen throughout the duration of the trial. After a short delay, an image of a tree was shown, together with the response contingencies for that trial. Participants could either "accept" or "reject" an offer to plant the displayed tree in this garden. The chosen response was highlighted. After a brief delay, numerical feedback was shown for the chosen, as well as the unchosen option. Rejecting a tree was always associated with a reward of zero. Accepting a tree yielded a reward/penalty that depended on the context and feature value (see A). (**D**) Training curricula. Two groups of participants were trained either on a blocked curriculum, in which the two contexts/gardens were blocked, or in an interleaved curriculum where the two gardens were randomly interspersed. All participants were subsequently evaluated on a randomly interleaved test phase without feedback.

Similar to this trees task, the neural network simulations described here involved binary categorisation of stimuli according to one of two task rules, which are defined by orthogonal category boundaries in feature space. In all simulations, the rules are explicitly cued by a contextual signal (which we also refer to as "task signal"), and fully supervised feedback is provided based on the context (task) and stimuli [59]. Thus, one can conceive the model as performing a task in which trees are categorised by leaf and branch density, or apples by size and colour. In practice, network inputs were simplified images of Gaussian "blobs", with the two relevant dimensions being the location of the peak along the x-and y-axis respectively (**Fig. 2a**). This allowed us a testbed that matched our domain of interest (e.g., inputs were high-dimensional, but two cardinal dimensions were relevant) without the potential biases that arise from naturalistic stimuli. We refer to the two "tasks" performed by the neural network as discriminating the peak of the blob with respect to lines that bisected the horizontal and vertical midlines respectively (**Fig. 2a**). We achieve very similar results using a reduced version of the trees task, although this requires a slightly more complicated neural network architecture; this is reported in the supplementary materials (Fig. S5-S7).

**Blocked vs interleaved training with standard SGD**

We began by training and evaluating a model we call the "vanilla SGD" network. The model is a fully connected feedforward network (multi-layer perceptron or MLP) with a single hidden layer, Rectified Linear Unit (ReLU) non-linearities and a single output node. We initialized the network with small random weights (σ=0.001), placing the network in the "rich" learning regime [15]. Inputs to the network were flattened images of Gaussian blobs, together with a one-hot encoded contextual cue signal (e.g. [0 1] for task 1 and [1 0] for task 2; see **Fig. 2b**. The network is trained using stochastic gradient descent (SGD) either on blocked data, where it is exposed to one task at a time over a prolonged training block, or on interleaved data where trials from both tasks are randomly interspersed within a single block (**Fig. 2c**). It is then evaluated on both tasks without supervision (i.e., with no further optimisation).

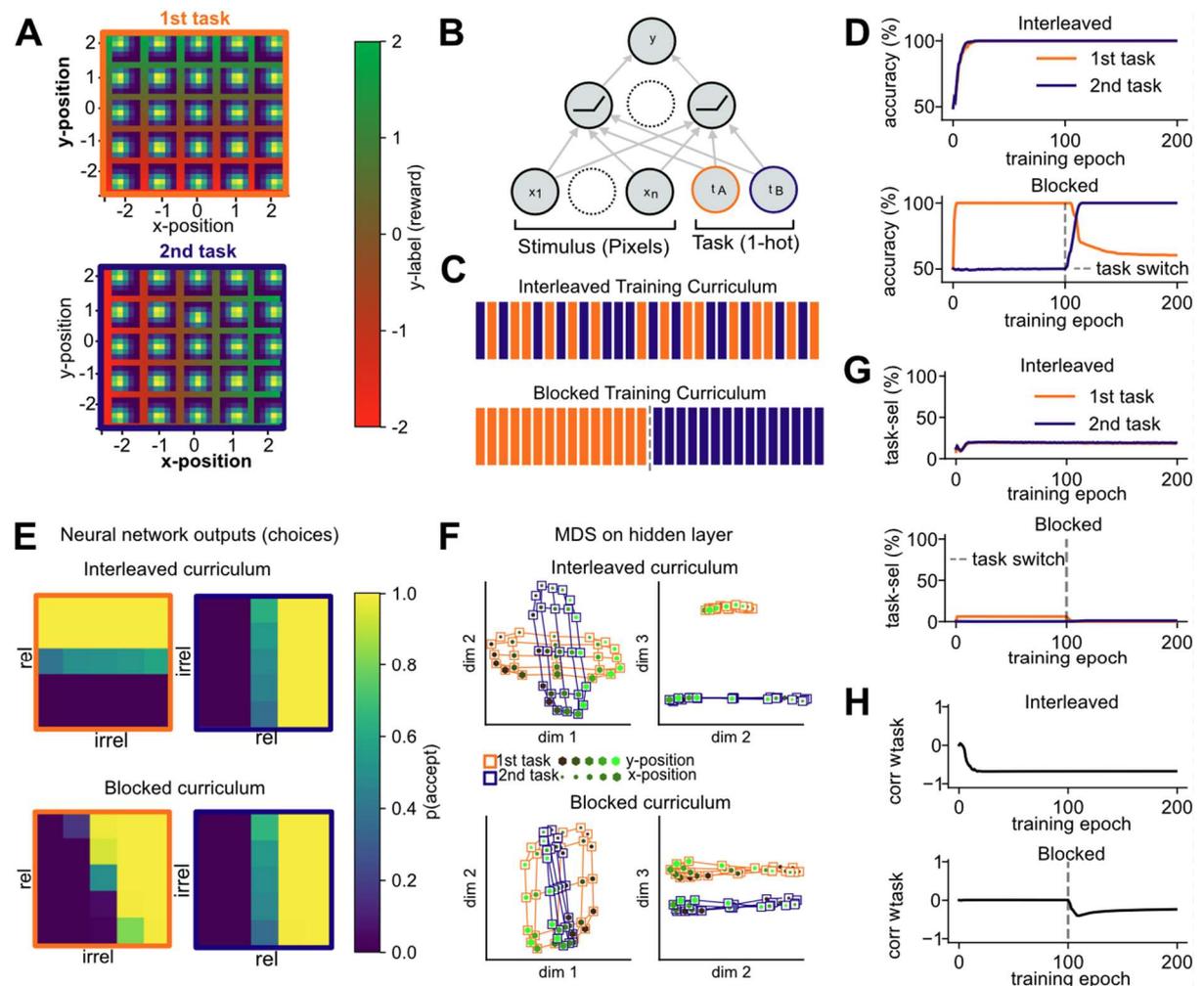

**Figure 2. Task design, network architecture and results of blocked versus interleaved learning. (A)** Task design. Stimuli were two-dimensional Gaussian functions ("blobs") for which we systematically varied the location of its peak along the x- and y- dimensions in five

discrete steps. Each subpanel visualises the Gaussian blob input image at that location in the underlying 2D stimulus space. Only one of the two feature dimensions was relevant per task, so that the reward (y-label) depended on the x-position in the first task (orange) and y-position in the second task (blue). **(B)** The network was a simple feed-forward MLP with a single hidden layer with ReLU non-linearities and received the flattened images of Gaussian blobs together with a one-hot encoded task signal as inputs. **(C)** The network was trained either in a fully interleaved curriculum in which trials from both contexts were randomly interspersed, or in a blocked curriculum in which it was first trained on one task, and then on the other. **(D)** Under interleaved training, the network quickly reached 100% training accuracy on both tasks. In contrast, under blocked training, learning the second task came at the cost of forgetting how to perform the first task. **(E)** Plotting the choices of the trained network in two dimensions revealed that under interleaved training, choices were aligned with the ground truth category boundaries (shown in (A)), whereas under blocked training, the network treated the first task as if it was the second. **(F)** Projections of the hidden layer activity into three dimensions via multi-dimensional scaling (MDS) shows orthogonal representations under interleaved training, where irrelevant information was suppressed, and parallel representations under blocked training, where the first task is encoded in the same way as the second task. **(G)** Under interleaved training, a significant proportion of hidden units were exclusively selective to the relevant dimension in one task (but not the other), whereas no such task-selectivity was observed under blocked training. **(H)** Evolution of correlation between task weights for both tasks during training. Interleaved - but not blocked - training promoted learning of anti-correlated task weights.

As expected, the vanilla SGD network suffered catastrophic interference when trained on each task in succession, with the ability to perform the first task overwritten by training on the second (**Fig. 2d**). Plotting network choices made during validation as a function of the two feature values (x- and y-location) revealed that the network applied the category boundary of the second task to the first task, ignoring the task signal (**Fig. 2e**). However, under interleaved training, the network converged to perfect performance, learning two orthogonal category boundaries, one per task. Projecting the hidden layer representations observed during validation into two dimensions confirmed that this network had learned task-specific manifolds under interleaved training. Each task was represented by a single axis that only encodes task-relevant information – the location along the x- or y-axis respectively. The axes were orthogonal to each other and separated by context along the third direction (**Fig 2f, upper**), a finding we had already observed in a previous study [15]. In contrast, after blocked training, the network represented the first task as if it were the second, and no longer distinguished between tasks (**Fig. 2f, lower**).

How did the network learn this representation? Previous work suggested that the pattern observed under interleaved training can be obtained via non-linear gating, if the context signal acts as additive bias to filter out irrelevant dimensions via context-dependent deactivation of units that encode task-irrelevant information [15]. In fact, 20% of units in the hidden layer became task-selective under interleaved training, responding to the relevant (but not irrelevant) dimension in one task and being active in the other task (**Fig. 2g, upper**). Under blocked learning, however, no such task-selective units emerged, suggesting that the network ignored the task signal (**Fig 2g, lower**). We have previously observed that the weights from the task units to the hidden units become anti-correlated over the course of interleaved training, pushing the input to the ReLU to positive or negative values depending on the context [15]. For the current simulations, this effect is shown in **Fig. 2h**. Under blocked learning, this anti-correlation does not emerge, as the network fails to utilise the task signal (**Fig. 2h**).

Taken together, thus, we found that in the vanilla SGD network, the two tasks were represented by allocating them independent hidden layer units, using context-dependent gating. This replicates our earlier report [15]. Under blocked training, the network failed to utilise the task units to implement this gating scheme, as the task signal was not required to solve individual tasks in isolation.

**Modelling the cost of interleaving with "sluggish" neurons**

During validation, humans are less accurate after interleaved compared to blocked training on the visual categorisation task [6,25]. In other words, they seem to show opposite behaviour to the vanilla SGD network, which had lower performance on blocked compared to interleaved training. We thus sought to develop a theory that could account for these discrepancies and devise algorithmic motifs that would more closely mimic those performance differences observed in human participants. How does this cost of interleaved training arise? In the real world, contexts tend to be temporally autocorrelated. Humans spend prolonged periods of time in one context, and switches occur intermittently (for example, when you leave the office to head home for the day, or when you leave the motorway and drive through an urban area). One possibility, thus, is that participants have an inductive bias that tasks should remain the same over time, in which case it is rational to condition behaviour not just on current task cues, but those that occurred in the immediate past [57]. This explanation has been offered for the ubiquitous observation that people are biased by the cues and responses that occurred on previous trials, and that switching between tasks incurs a cost to accuracy and RT [60]. Here, we propose that in humans, these choice history biases create interference during interleaved, but not during blocked learning (see [54] for a related account). Previously, we hypothesised that this may lead humans to ignore the context signal and effectively apply the same categorisation rule irrespective of the context, which optimises for performance on congruent trials (those with the same responses across tasks.) (**Fig. 3a, lower**) [25]. In contrast to this linear solution, with blocked training, human participants can effectively factorise the decision problem and learn one rule per task (**Fig. 3a, upper**).

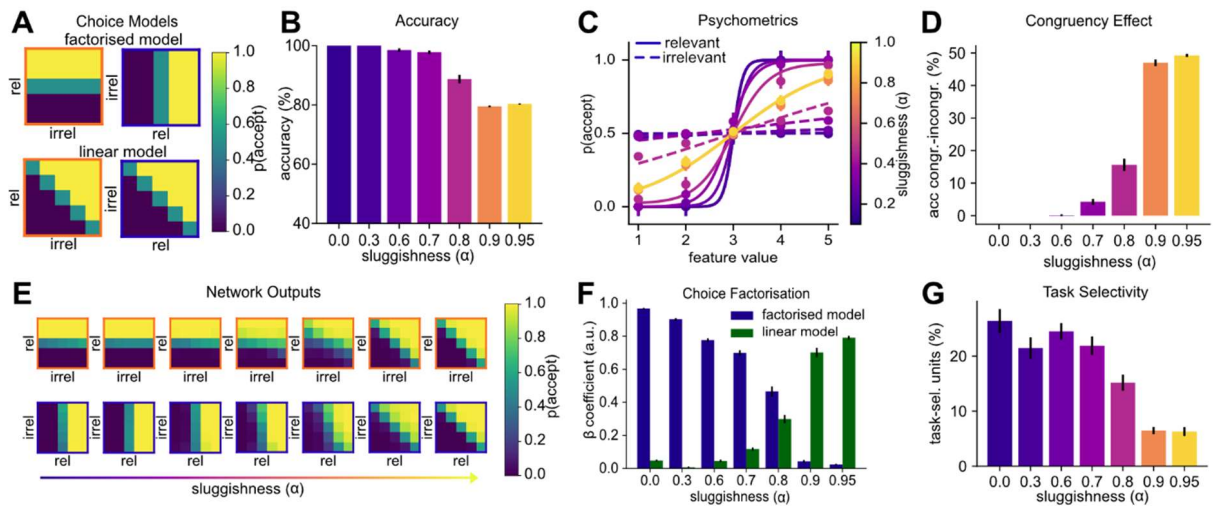

**Figure 3. Modelling the cost of interleaving with a "sluggish" task signal. (A)** Illustration of the cost of interleaved training. The factorised model (top) assumes that two separate category boundaries are learned, one for each task. The linear model (bottom) assumes that the task signal is ignored, leading to the acquisition of a diagonal category boundary that yields high performance on both tasks. We hypothesised that interleaved training would promote a solution as predicted by the linear model. **(B)** Test phase accuracy of neural networks trained on interleaved data with different levels of "sluggishness" (exponential average of the task signal). The higher the sluggishness, the lower the task accuracy. **(C)** Sigmoidal curves fit to the choices of networks described in (B). The solid lines indicate how the choices depend on the relevant dimension and the dashed line how they depend on the irrelevant feature dimension. As the sluggishness increases, sensitivity to the relevant dimension decreases and to the irrelevant dimensions increases. **(D)** Difference in accuracy between congruent and incongruent trials (i.e., those with the same or different responses across tasks). The congruency effect depends on the amount of sluggishness. **(E)** Network outputs (choices) for different levels of sluggishness. As sluggishness increases, the networks move from learning a "factorised" to learning a "linear" solution. **(F)** Linear regression coefficients obtained from regressing the outputs shown in (E) against the models shown in (A), confirming that sluggishness controls whether a factorised or linear solution is learned. **(G)** Proportion of units in the hidden layer which are task selective. With increasing sluggishness, fewer units are exclusively selective for one task.

To model this cost and tendency towards a linear solution, we introduce the concept of "sluggish" units, that is, neurons that carry information from previous trials over to the current trial [58]. We model this sluggishness with an exponentially moving average (EMA, see methods) with the weight on previous trials controlled by a single parameter, $\alpha$. Setting $\alpha = 0$, is equivalent to the vanilla SGD network described above; other models are "sluggish SGD" networks. Increasing $\alpha$ has the effect of decreasing performance at validation overall (**Fig. 3b**). In **Fig. 3c**, we plot psychometric data, i.e., the effect of $\alpha$ on how response probability varies with relevant and irrelevant information. Visual inspection suggests that the parameter controls the extent to which information along the irrelevant dimension is factored into the model's choices (**Fig. 3c**).

Plotting the choices in two dimensions offers further insights into the effect of sluggishness. As $\alpha$ increases, the model moves from learning a factorised solution with one boundary per task to a linear solution with a single category boundary (**Fig. 3e**). Indeed, the factorised model fit better for low sluggishness values, whereas the linear model fit better for larger sluggishness values (**Fig. 3f**). In other words, the sluggishness introduces a congruency effect, whereby the network performs much better on trials with the same label across tasks (congruent) compared to trials with task-unique labels (incongruent) (**Fig. 3d**). At the level of neural representation, we observed a reduction of the proportion of task-selective hidden layer units (with axis aligned tuning profile) relative to task-agnostic units (selective for congruent trials) (**Fig. 3g**).

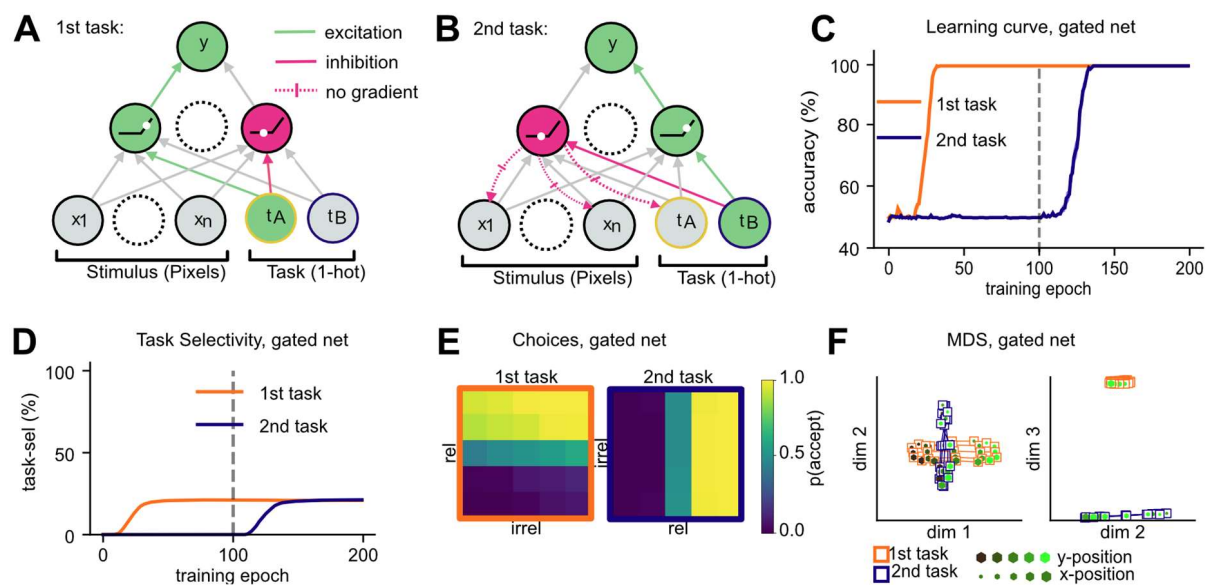

**Figure 4. Blocked training with manual context gating. (A)** Illustration of how task weights can be set to values with opposing signs to activate and deactivate units in the hidden layer while the network is trained on the first task. **(B)** Same as (A), but for the second task. Manually settings task weights to anti-correlated values (different signs for first and second task) ensures that different units are active in each task. **(C)** Learning curves of networks trained with manual gating, showing that forgetting has been mitigated. **(D)** As training progresses, task-selective units emerge for both tasks. Gating protects units selective for the first task during training on the second task. **(E)** Network outputs are axis-aligned, just as seen previously under interleaved training. **(F)** MDS on the hidden layer revealed orthogonal representations where task-irrelevant dimensions are attenuated.

**Modelling blocked learning with non-linear gating**

While the introduction of "sluggish" neurons imposes a cost on interleaved training, it doesn't solve the problem of catastrophic forgetting under blocked training. How can we account for the ability of humans to learn continuously without substantial forgetting? The vanilla SGD network trained on interleaved data learned a factorised representation where different populations of hidden units were allocated to the first and second task. This allocation was achieved via non-linear gating, implemented by the task weights which connected the task-signalling units to the hidden layer and pushed the hidden layer activity into the negative/positive input range of the ReLU non-linearities. We wondered whether this simple gating mechanism that allocates different subsets of units to different tasks may be sufficient to guard against forgetting. To test this, we first hand-crafted the gating scheme by manually setting the weights that connect task units to hidden units to anti-correlated values, such that each unit received a positive bias in one task and a negative bias in the other (**Fig. 4a,b**). We then trained the remaining units end-to-end on a blocked curriculum. This network no longer forgot how to perform the first task after it was trained on the second (**Fig 4c**), which suggests that a simple gating intervention that partitions the hidden layer may be sufficient to guard against catastrophic interference. The outputs of the network were axis-aligned, demonstrating that it learned accurate representations of the two category boundaries (**Fig. 4e**). At the level of hidden units, we observed once again orthogonal and low-dimensional manifolds that encoded task-relevant and suppressed task-irrelevant dimensions in a context-dependent manner, just like in the vanilla SGD network trained on interleaved data (**Fig. 4f**). Note that [54] describes a closely-related set of simulations and equivalent results in this handcrafted setting, and the result described here is consistent with a previous literature proposing gating as a solution to continual learning [38,51].

**Anti-correlated task weights via Hebbian learning**

Ideally, we would like these gating signals to be acquired without intervention by the experimenter. Thus, we introduced another algorithmic motif: the use of a Hebbian learning step following supervision. Due to the one-hot representation of the context variable, the context units are correlated with those hidden units that encode task-relevant information for the active context. Consequently, the Hebbian step strengthens the connections between the task context units and those hidden units encoding task-relevant information and weakens the connection to units coding for irrelevant information. We use a variant of Hebbian learning with weight-decay, called Oja's rule [61,62]. A well-known property of Oja's rule is that it converges to the first principal component of the inputs when applied to mean-centred data. Crucially, in our simple case of only two tasks, the direction of largest variance in the mean-centred input space of our Gaussian blob dataset is spanned by the two task signals (**Fig. 5a, b**). Indeed, when performing weight updates with Oja's rule on a single hidden unit, that unit recovered the first principal component of the input dataset, which distinguished between the two contexts. We observed that the two weights between the context units and the hidden unit converged to values with opposing signs, the desired requirement for non-linear gating (**Fig. 5c**).

We concluded that Hebbian updates with Oja's rule could be used to establish links between the task signal units and active units in the hidden layer. To implement this, we extended this approach to multiple hidden units, so that each of these would learn to receive task signals via anti-correlated weights. As for the handcrafted solution in **Fig. 4**, when stimuli were propagated forward through the network to the hidden layer, those units that had positive outputs for task A had negative outputs for task B and vice versa. Thus, applying a ReLU nonlinearity partitions a portion of the hidden layer into task A and task B selective units. To assess whether this Hebbian learning step would be sufficient to guard against catastrophic forgetting, we devised a new training scheme in which we alternated the supervised SGD update and the Hebbian update on each training step (methods). We call this model the "Hebbian Gating" network. Crucially, we found that this intervention was sufficient to alleviate catastrophic forgetting. The performance of the network on the first task remained at ceiling, even after training on the second task (**Fig. 5d**). Just as in the vanilla SGD network trained on interleaved data, we observed that for the Hebbian Gating network the learned task weights were anti-correlated even for blocked training (**Fig. 5f**). Thus, the hidden layer was partitioned into task A and task B selective units (**Fig. 5e**) and the representations embedded in the hidden layer population response became orthogonal, with compression along the irrelevant dimensions (**Fig. 5h**), a factorisation that was also reflected in two accurate category boundaries at the output level (**Fig. 5g**).

We note that in practice, the solution is somewhat sensitive to the length of the training block and requires a carefully tuned balance between the learning rates for the supervised and Hebbian updates. When we repeated the simulations and systematically increased the length of the training blocks, whilst keeping all other parameters constant, the network forgot more about the previous task the longer the training blocks were (**Fig. S1a**). However, even when we doubled the block length, its performance was still superior to the baseline model trained without the Hebbian updates (**Fig. S1b**).

To summarise, we have demonstrated how a variant of Hebbian learning can be used to learn anti-correlated weights that connect task units to relevant hidden units, and that alternating between supervised and Hebbian training updates allows a network trained on blocked data to learn tasks sequentially without forgetting. Representations formed by the network were identical to those observed under interleaved training in the vanilla SGD network.

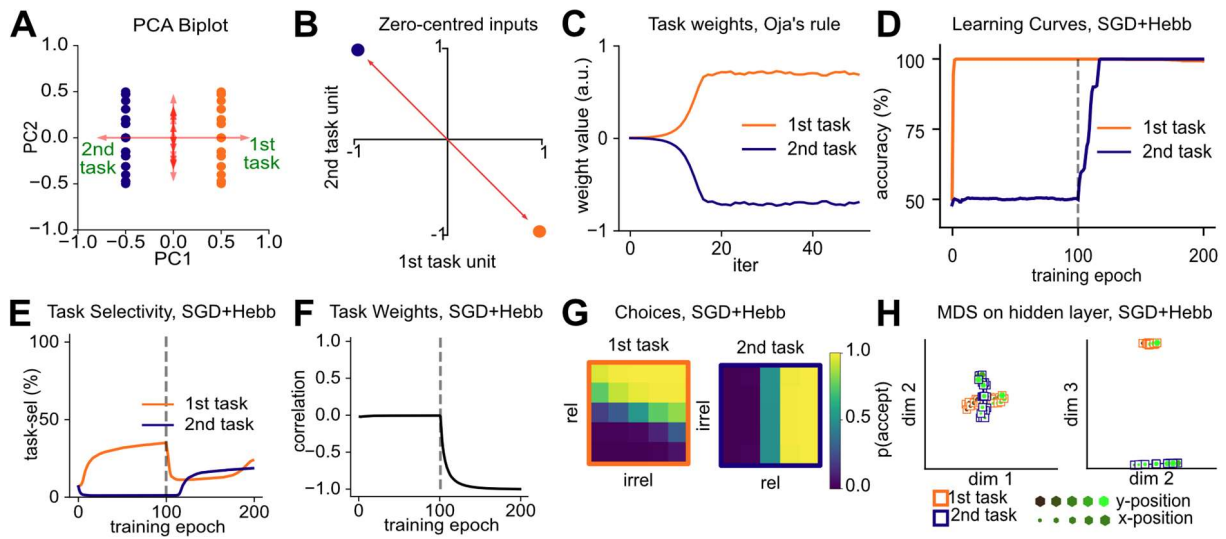

**Figure 5. Protection against catastrophic forgetting via Hebbian learning. (A)** Biplot on training trials. The axis of largest variation is spanned by the context signal. **(B)** Schematic showing why the first component is spanned by the context signal (for zero-centred inputs). **(C)** Weights from task units to a single hidden unit, trained with Oja's rule. Over time, weights with opposing signs emerge. **(D)** Learning curve of neural networks trained with SGD and Oja's rule in alternation. The network no longer forgets how to perform the first task. **(E)** Task selectivity of hidden layer units. SGD+Hebbian learning prevents the network from losing units that are selective to the first task. **(F)** Hebbian learning induces the desired anti-correlation between task weights. **(G)** Network outputs are axis aligned under blocked learning, just as previously seen under interleaved learning. **(H)** MDS on the hidden layer, revealing orthogonal representations that encode both tasks without interference.

**Modelling human continual learning with Hebbian context gating**

Next, we assessed whether our two algorithmic innovations, the sluggishness and the Hebbian update step, were sufficient to reproduce error patterns made by human participants who had been trained on a comparable task. We re-analysed a dataset from a previous study in which participants learned to accept/reject images of fractal tree stimuli in two different task contexts, introduced as the north and south garden [25]. Just as for our Gaussian blobs, trees varied along two different feature dimensions, corresponding to the density of leaves ("leafiness") and number of branches ("branchiness"), of which only a single dimension was relevant for each task. The participants were trained either on a blocked curriculum, or on a randomly interleaved curriculum. Crucially, participants whose training phase was blocked performed better at a subsequent interleaved validation phase, compared to those who received an interleaved training curriculum. Further analyses of the error patterns revealed that these participants had better estimates of the decision boundaries for each task and were less influenced by variation along the task-irrelevant dimensions. To assess the effectiveness of our approach, we compared validation performance after blocked or interleaved training between a neural network with both innovations, the sluggishness and the Hebbian updates (called "sluggish Hebbian gating network"), and a standard feed-forward neural network that was trained without any further algorithmic innovations ("vanilla SGD network"). To perform statistical inference on the neural networks, we collected 50 independent training runs with randomly initialised networks per training curriculum. We adapted the learning rates of both networks to make it possible to learn with the same number of trials as the human participants in the previous publication (200 trials per task). Even with only such a small number of training trials, the networks replicated all key observations reported earlier (**Fig. S2**). Moreover, in contrast to the baseline MLP, the network equipped with sluggishness and Hebbian update step qualitatively recreated all key aspects of the human behavioural data.

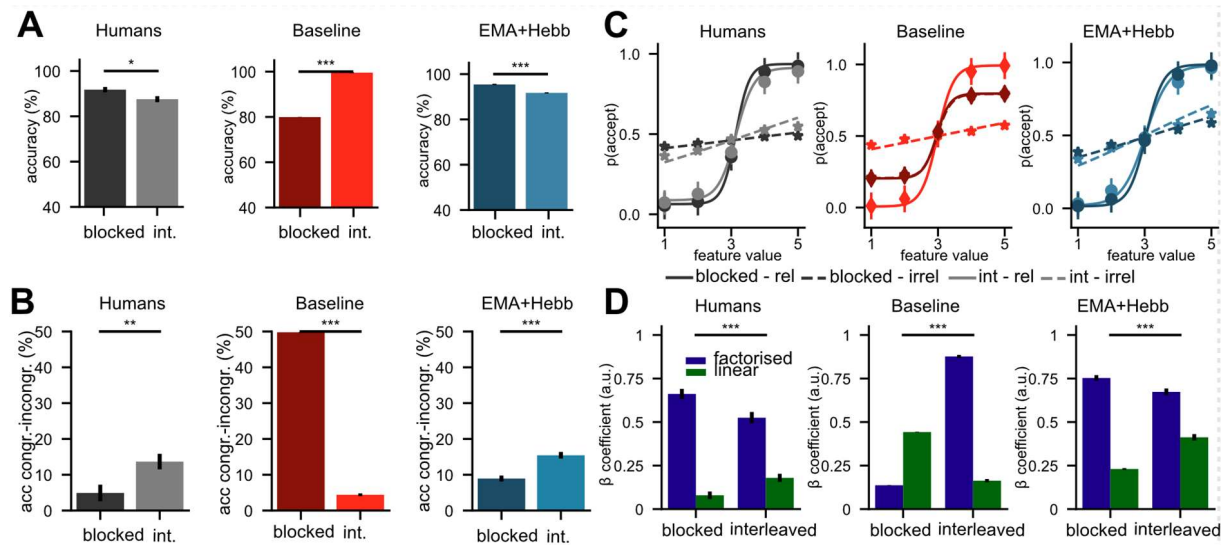

**Figure 6. Modelling the benefit of blocked over interleaved training observed in data from human participants. (A)** Test phase accuracy. Humans perform better under blocked compared to interleaved training, while the baseline model performs better under interleaved training, due to catastrophic forgetting. Our Hebbian network with sluggish task units performs better under blocked training. **(B)** Congruency effect. Just like human participants, under interleaved training, our model performs worse on incongruent compared to congruent trials. **(C)** Sigmoidal fits of choices made by human participants, the baseline network, and our network. Our network recreates the intrusion of irrelevant dimensions observed under interleaved, but less so under blocked training. **(D)** Fits of the factorised and linear model to human and network choices. In contrast to the baseline model, our model recreates patterns observed in humans, where the factorised model fits better under blocked than interleaved training.

First, human participants trained on a blocked curriculum had a higher test accuracy than those trained on interleaved data (*T(93)=2.32, p=0.022*, **Fig. 6a, left panel**). While the opposite was true for the vanilla SGD network, which suffered from catastrophic interference (*T(98)=-95.94, p<0.0001*, **Fig. 6a, middle panel**), the sluggish Hebbian Gating network showed a similar benefit of blocked over interleaved training at test (*T(98)=5.71, p<0.0001*, **Fig. 6a, right panel**). Our modelling of the impact of sluggishness on task performance revealed a congruency effect: The "sluggish" network performed better on congruent than incongruent trials. Hence, we wondered whether participants showed a similar congruency effect, and whether this difference would be larger in the interleaved group, where participants tended to use the same decision boundary for both tasks. Indeed, human participants showed a strong interaction between the training curriculum and the congruency effect, which was larger under interleaved training (congruency blocked vs interleaved: *T(93)=-2.74, p=0.007,* **Fig. 6b, left panel**). Due to catastrophic forgetting, the congruency effect was much larger under blocked training in the vanilla SGD network (*T(98)=112.07, p<0.0001,* **Fig. 6b, middle panel**), while our novel training procedure for the sluggish Hebbian Gating network recreated the effect observed in humans (*T(98)=-5.07, p<0.0001,* **Fig. 6b, right panel**). Next, we fitted psychometric functions (sigmoid) to the choices made by human participants and by our models, separately for the relevant and irrelevant feature dimensions. In humans, slopes for the irrelevant dimension were significantly steeper under interleaved than blocked training, suggesting that choices of these participants were stronger influenced by task-irrelevant information (blocked vs interleaved: *T(93)=-2.77, p=0.0068*, **Fig. 6c, left panel**). Choices made by the vanilla SGD network followed the opposite pattern, with more intrusions from irrelevant dimensions under blocked training (blocked vs interleaved: *T(98)=82.99, p<0.0001,* **Fig. 6c, middle panel**). In contrast, the sluggish Hebbian Gating network was less influenced by irrelevant feature dimensions under blocked compared to interleaved training (blocked vs interleaved: *T(98)=-7.32, p<0.0001*, **Fig. 6c, right panel**).

How did participants learn the two tasks? The original paper suggested that human participants learned "factorised" representations under blocked, but less so under interleaved training. To test this, we fit the factorised and linear model described earlier to the choices made by the models. For human participants, the factorised model explained choices better under blocked than under interleaved training (*T(93)=3.07, p=0.0028*, **Fig. 6d, left panel**), while the opposite was true for the linear model (*T(93)=-3.12, p=0.0024,* **Fig. 6d, left panel**). As expected, the opposite patterns were observed for the vanilla SGD network (*T(98)=-79.72, p<0.0001*, *T(98)=27.98, p<0.0001,* **Fig. 6d, middle panel**), which learned to factorise the problem under interleaved, but not blocked training. The sluggish Hebbian Gating network recreated the patterns observed in humans, suggesting that it learned two accurate decision boundaries under blocked, but not under interleaved training (*T(98)=3.03, p=0.0044*, *T(98)=-9.30, p<0.0001*, **Fig. 6d, right panel**).

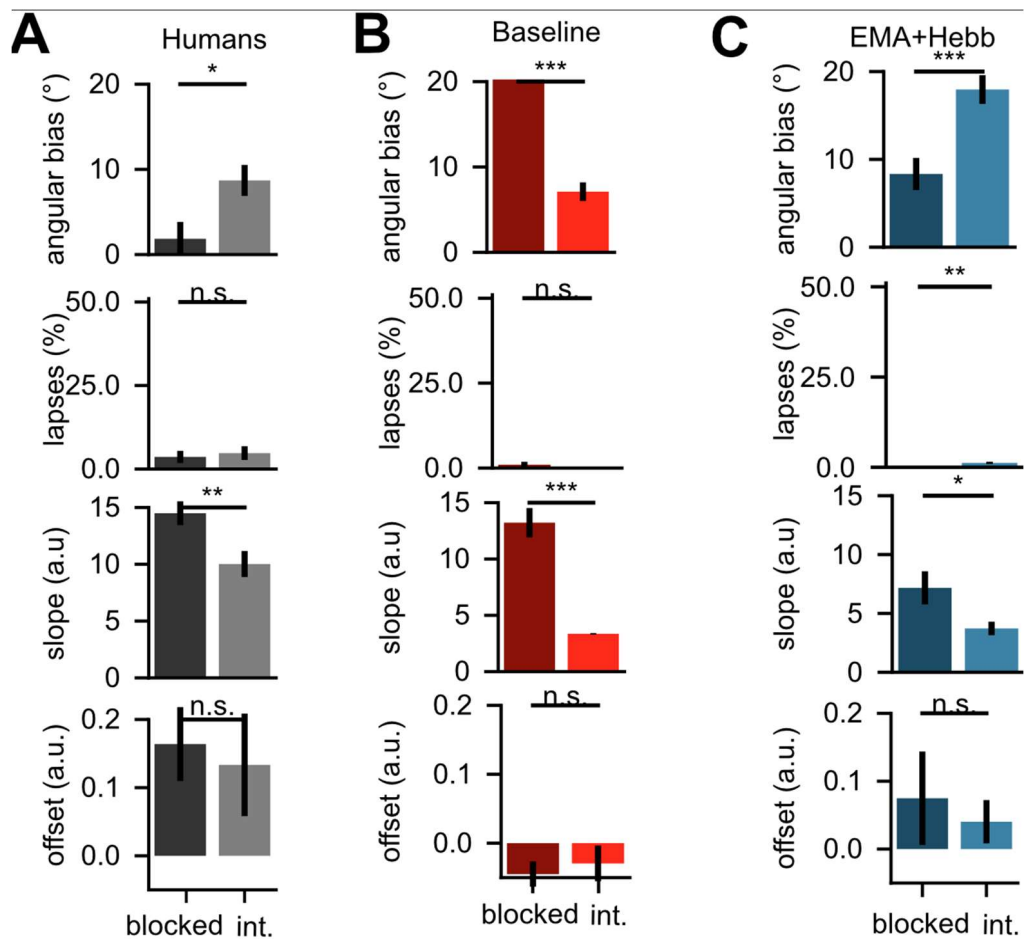

**Figure 7. Fits of Psychometric model to human behaviour and neural networks. (A)** Fits of a psychometric model with four parameters to choices made by human participants, decomposing error patterns into (i) angular bias of category boundary (ii) lapse rate (iii) slope and (iv) offset of sigmoidal transducer. Both the angular bias and slope differed significantly between participants trained on blocked and interleaved curricula. **(B)** Same as (A) but fitted to the vanilla SGD network. **(C)** Same as (A), but fitted to the sluggish Hebbian Gating network.

However, intrusions from the irrelevant dimensions might not have been the only source of errors. It was also possible that one group made more unspecific errors (lapses), was less sensitive to information along the relevant dimension or exhibited a systematic bias in the offset of their learned category boundary. Using a psychophysical model with free parameters for the angle of the learned category boundary, the number of unspecific errors, the slope and offset of the sigmoidal transducer showed that the length of training blocks predominantly affected the accuracy of the category boundary estimate [25]. Our reanalysis of the human behavioural data confirmed this, with larger angular biases in the interleaved compared to the blocked group and a significant difference in slope, while differences in lapse and offset parameters were non-significant (*bias: T(93)=-2.54, p=0.0127, lapse: T(93)=-0.41, p=0.6807, slope: T(93)=2.88, p=0.0049, offset: T(93)=0.33, p=0.7419*, **Fig. 7a**). The vanilla SGD network had a significantly larger angular bias in the blocked group, due to catastrophic forgetting of the first task (*bias: T(98)=16.51, p=0.0000, lapse: T(98)=0.73, p=0.4721, slope: T(98)=7.57,*

*p<0.0001, offset: T(98)=-0.49, p=0.6260,* **Fig. 7b**). In contrast, fits to the sluggish Hebbian Gating network were similar to those observed in human data (*bias: T(98)=-3.91, p=0.0004, lapse: T(98)=-3.44, p=0.0014, slope: T(98)=2.27, p=0.0289, offset: T(98)=0.46, p=0.6512,* **Fig. 7c**). Taken together, these findings demonstrate how two adjustments to the training procedure, the introduction of sluggish task signals and a Hebbian learning step that is alternated with SGD updates, are sufficient to protect against catastrophic forgetting and model the cost of interleaved training observed in human participants.

**Very sluggish task estimates under interleaved training bias internal representations**

Why did the "sluggish" task signal lead to intrusions from irrelevant dimensions? In the original paper, we hypothesised that humans benefit from blocked training as it aids the formation of "factorised" representations, while interleaved learning might induce shared representations [25]. In subsequent neuroimaging work, we found evidence for such factorised and orthogonal representations in fronto-parietal areas of the human brain after blocked training [15]. However, it is less clear how interleaved training might shape internal representations. We hypothesised that while blocked training with Hebbian updates should lead to orthogonal representations, interleaved training might induce representations that preferentially encode congruent stimuli, i.e., those that required the same response across tasks and lie on the main diagonal of the two-dimensional stimulus space.

To test this, we regressed RDMs from the hidden layer of our models trained with large sluggishness values and either on blocked or interleaved curricula against a set of candidate RDMs encoding grid-like, "orthogonal", or "diagonal" representations. The grid model served as control and assumed that both feature dimensions were encoded in both tasks, forming a task-agnostic representation. In contrast, the orthogonal model represented the case where, starting from this grid model, task-irrelevant feature dimensions were filtered out, leaving a task-specific representation that encodes the relevant dimension in each context, with the two representations being orthogonal to each other. Lastly, in the diagonal model, representations of the stimuli were projected onto the main diagonal of the two-dimensional stimulus space which corresponded to stimuli that required the same response across tasks (methods). Indeed, while the orthogonal model explained the patterns best under blocked training *(grid vs orthogonal: T(49)=-347.62, p<0.0001; orthogonal vs diagonal: T(49)=178.67, p<0.0001)*, the diagonal model, which represented congruent stimuli, was the best predictor of hidden layer activity under interleaved training (*grid vs diagonal: T(49)=-109.23, p<0.0001; orthogonal vs diagonal: T(49)=-73.89, p<0.0001*, **Fig. 8a**). How were these representations formed? Assessing the task-selectivity of individual units in the hidden layer revealed that while a sizeable fraction of units was selective to the relevant dimensions of each task under blocked learning (41.3%), most hidden units of the network trained on interleaved data were task-agnostic (99.4 %, **Fig. 8b**). Lastly, in the model trained on an interleaved curriculum, readout weights from those task-agnostic units were significantly larger than those reading out from the task-selective weights (*task agnostic vs 1$^{st}$ task: T(49)=5.52, p<0.0001; task agnostic vs 2$^{nd}$ task: T(49)=5.58, p<0.0001*, **Fig. 8c**). Together, these analyses suggest that interleaved training might not only alter the readout, but also the geometry of task-representations, providing avenues for further empirical research.

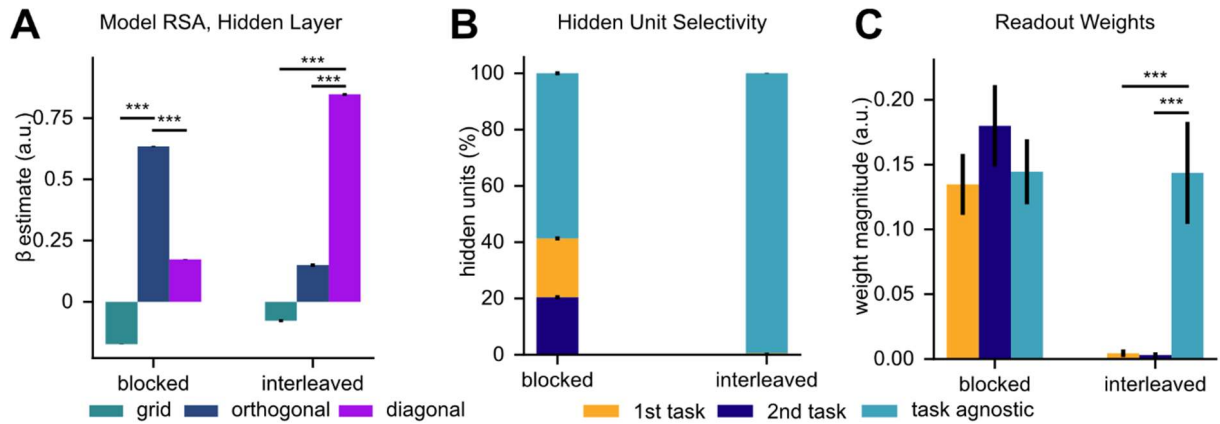

**Figure 8. Very sluggish task estimates under interleaved training bias internal representations. (A)** Coefficients obtained by regressing hidden layer RDMs against three model RDMs, encoding grid, orthogonal or diagonal representations. The orthogonal model fits best under blocked training, whereas the diagonal model fits best under interleaved learning, suggesting that the latter aligns the representations with stimuli encountered in congruent trials. **(B)** Proportion of task selective units under both curricula. Under blocked learning, more units are task-selective than under interleaved learning. **(C)** Magnitude of readout weights, showing that under interleaved learning, the network relies more strongly on task-agnostic units (which encode congruent trials, see also **Fig. S3**).

## DISCUSSION

Previous work has shown that humans perform worse after blocked compared to interleaved training on multiple categorisation tasks [6,25]. In contrast, to converge, deep neural networks require training data to be randomly interleaved, as they suffer from catastrophic forgetting under blocked curricula. This limits both their performance and their viability as a model of human learning [1,25]. Here, we propose a neural network model of human continual learning which captures this benefit of blocked over interleaved training and recreates several observations made in human participants at the behavioural and neural level. First, we demonstrated how a "sluggish" task signal introduces biases in the acquired task representations which leads to worse performance under interleaved training. We note earlier reports that have previously proposed similar approaches to account for the cost of interleaving [54,58]. Secondly, we showed how gating, an inherent property of prefrontal cortex function, could not only be used to control switching between already learned tasks, but might indeed play an active role in the acquisition of novel tasks without forgetting. We propose that by augmenting standard supervised training with a Hebbian update, this gating scheme can be learned from scratch. Building directly on previous work on representation learning in humans and neural networks, we illustrated how these two properties shape neural representations, and how the emerging representational geometry can influence behaviour. Lastly, we validated our model by fitting it to previously published human behavioural data, allowing us to recreate the performance difference between blocked and interleaved training. Decomposition of these differences into different sources of error revealed that in both human participants and our model, differences were predominantly driven by a misestimation of the category boundary under interleaved training.

The idea that sluggish neurons could model costs associated with task switching is not new. In early models of cognitive control, it was assumed that the PFC has a bias to maintain task information over time [63]. This "active maintenance" of task information would lead to intrusions between competing objectives and could explain why humans usually perform worse immediately after a switch to a different task [64]. Here, we extended this idea and investigated how switch costs shape credit assignment during learning, demonstrating that interleaved training impairs the ability to link relevant perceptual information to the correct contextual cue.

A key component of our model is non-linear gating of internal representations. Early connectionist models have demonstrated how gating could be utilised by PFC to minimise interference during multi-tasking [43,49], and follow-up work suggested that basal ganglia could control the gating of PFC representations [63]. However, with few exceptions such as [44] and [15], the gating was usually hand-crafted by the experimenters, and it remained unclear how these control processes might emerge in the first place. Similarly, a handful of studies have drawn the link to continual learning and investigated how gating could prevent catastrophic forgetting, but once again, the process was usually implemented by hand [38,51,54]. We demonstrate that a simple biologically inspired intervention (Hebbian learning) is sufficient to implement this gating strategy. At a representational level, the gating effectively orthogonalises hidden layer representations by enforcing an axis-aligned coding scheme. Interestingly, a recent series of papers has provided converging evidence that the brain might use orthogonal representations to minimise interference between tasks [15,41,42,65] and some of the more successful recent engineering solutions to Catastrophic Forgetting employ orthogonalization of gradient updates of internal representations [31–33]. Here, we propose a biologically inspired model of how these orthogonal representations could be learned.

A possible limitation of our approach is that it requires that the largest principal component of the input space is the task/context signal. We note that our solution was designed to meet two objectives: First, to identify the context signal among all inputs, and secondly to use this signal for context-dependent gating. In the task we studied, the Hebbian update solved both problems, as it identified the largest PC in the dataset, which happened to be the context, and linked it to task-selective hidden units. An alternative formulation of the problem would leave the context discovery to another mechanism. If we had already identified the source of the context signal, the Hebbian step could still be used to learn the gating procedure. To demonstrate this, we ran a complementary simulation in which we applied Oja's rule exclusively to the weights from the two task-signalling units to the input of the hidden layer. In this case, our mechanism still learned to gate out task-irrelevant information (**Fig. S4**). This begs the question of how the context could be identified if it was not the largest principle component. One possibility is that attentional mechanisms serve as gain modulation that scales up neural activity coding for parts of the visual input that represents the context.

Our model appears to be strongly related to a recently published conference submission in which the authors demonstrated that carrying over task-signals from previous trials leads to lower performance on interleaved curricula [54]. Like in our work, the authors propose an implementation of "sluggishness" that is inspired by models of switch costs in PFC and suggest a simple gating mechanism to prevent forgetting. However, while the authors implemented this gating scheme manually, we propose a Hebbian training step that can learn this scheme from scratch. Leaving differences in implementational details aside, both studies provide converging evidence that theories on the role of PFC for cognitive control can be readily extended to the problem of continual learning.

How could the gating scheme be implemented in neural circuits? Our approach, motivated by the form of gating that is learned by a network with ReLU non-linearities in the interleaved setting, used a weighted additive input to the non-linearity. Prefrontal gating-like mechanisms have often been hypothesised to underlie context-dependent behaviour in humans and animals, including the gating of sensory information [66,67] and gating of task-relevant activity [45–47]. Such gating could be realised for example through top-down additive control [49]. However, a recent comparison of alternatives suggests that tuning this architecture, specifically using multiplicative forms of gating that act on the output of the non-linearity [38,63], might result in greater task accuracy and support generalisation across tasks [53]. Multiplicative gating could be implemented by neural oscillations [68], neurotransmitters [63,69] or even through dendritic properties of neurons [70,71].

There are several clear avenues for future research. We introduced the notion of sluggishness to account for performance costs observed in human participants under interleaved training. Similar to other recent accounts, we assumed this sluggishness to be an inherent property of prefrontal function [54]. Future work could investigate the normative basis of such a coding scheme. For example, under blocked training, the active maintenance of task signals might protect against noise in the task signal. Under this account, sluggishness would ensure ongoing task performance under blocked curricula, even if the task signal could not be read out or was mislabelled on a subset of trials. An even stronger claim, building on previous work on sequential effects in human decision making [57], would be that sluggishness might adapt to the volatility of the environment. Future work could investigate if the window over which contextual information is averaged depends on the amount of time spent in a single context, or the extent to which task switches are predictable from recent trial history.

The particular problem we studied provided only limited opportunity for cross-task transfer, and an optimal solution was to partition the network in separate sub-networks, one for each task, and humans appear to learn such a representation with blocked training curricula. It should be noted that blocked training is not always advantageous, as there seem to be several cases in which humans benefit indeed from interleaved curricula [72,73]. The extent to which either blocked or interleaved learning is advantageous might depend on the similarity between the to be learned tasks and hence the opportunity for cross-task transfer [21,74]. In our simulations, we observed that at the level of hidden units, task-selective units formed receptive fields that were aligned with the task-relevant dimensions, while task-agnostic units appeared to be selective for congruent trials, i.e., those that afford the same response across tasks (see also **Fig. S3**). Interestingly, sluggish task signals promote the formation of shared representations that don't arbitrate among tasks and read out from these task-agnostic units. A prediction that arises from these simulations is that sluggish units might help the learner to find similarities among tasks that are encountered in close temporal proximity. Consequently, it is likely that whether sluggishness introduces a cost or benefit for learning depends on the similarity between tasks and their transfer demands, and hence the need for shared or separated representations [3,75].

Another possible line of enquiry is lifelong learning. We focused on a simple and tractable context-dependent decision-making problem with only two tasks, using a small feed-forward neural network. Future work could investigate how this approach extends to additional tasks, both at the human behavioural level and in artificial neural networks. We note that our Hebbian procedure is in essence achieving a temporal clustering of contextual information, with the active cluster gating on a set of units and inhibiting the rest. This scheme in principle might work in richer settings with additional tasks. Real lifelong/continual learning, however, is likely to involve more than a single Hebbian learning mechanism applied to prefrontal gating signals. For example, a previous study that has investigated the utility of gating for continual learning at a scale noted that it was insufficient to protect against forgetting as the number of tasks increased [38]. By combining gating with a regularisation approach that prevented task-relevant weights from adapting to novel tasks, the authors were able to overcome this limitation. The importance of regularisation schemes was also noted in another recent paper that investigated the representations that emerge in neural networks trained on many cognitive tasks [16]. Future work could investigate how these gating processes interact with other extant solutions to catastrophic interference, such as memory consolidation and replay of previous experiences during sleep.

We focussed on a simple context-dependent decision-making problem that could be trivially solved with a standard feed-forward neural network with a single hidden layer. Future work could investigate how this solution scales to more complex datasets, tasks, and architectures. As a first step, we ran an additional simulation in which we trained a slightly deeper neural network with two hidden layers on the actual tree images from the original branch/leaf task. In this setup, context, signalled by one-hot units, was no longer the largest direction of variance, unless we multiplied it with a very large scalar weight. Even then, the training process remained quite unstable. To test whether our approach could in principle still work, we restricted the Hebbian updates to the weights from the two task units to the hidden layer, which enabled the network to learn both tasks continually without catastrophic forgetting and produced error patterns and hidden layer representations very similar to those we had observed with the "blobs" dataset and the smaller network (**Fig. S5**). Next, we introduced the sluggishness and performed a qualitative fit to the human behavioural data, which revealed that this architecture could still model the benefit of blocked over interleaved training we had previously observed in humans (**Fig. S6-7**). We take this to imply that the two algorithmic motifs, Hebbian learning for context-gating and sluggishness, can be applied to slightly more complex input datasets and bigger networks. However, gating strategies might be particularly suitable for context-dependent decision making with orthogonal rules, as the problems can be trivially solved by filtering out irrelevant dimensions. Related work has shown how regularisation approaches can be used to learn a much larger variety of cognitive tasks, such as delayed-match-to-sample and go/no-go paradigms [16]. Future studies could test how our approach extends to these paradigms and when it might break down. Another line of inquiry could explore sluggishness and Hebbian gating in more biologically plausible architectures that involve recurrence. Previous work has demonstrated that hand-crafted gating strategies [38] and weight orthogonalization procedures [34] can be adapted to RNNs. While the sluggishness could trivially be implemented in such an architecture, more work would be required to adapt the Hebbian update step.

To conclude, we introduced two algorithmic motifs to augment vanilla neural networks trained with stochastic gradient descent, "sluggish" task signals and a Hebbian update step, which together are sufficient to model the benefit of blocked over interleaved training previously observed in humans. Furthermore, investigation of the learned representations suggests that blocked training might promote the formation of orthogonal representations, like those observed in biological brains, while interleaved training leads to shared representations that optimise for congruent trials. Taken together, we provide a biologically inspired model of human continual learning, grounded in previous work on representation learning and the function of prefrontal cortex.

## MATERIALS & METHODS

**Software**

All simulations were implemented in Python 3.9 with the PyTorch 1.71 package. Hyperparameter tuning was carried out with the RayTune 1.10 package. Parallelisation was implemented with the joblib 1.0.1 package. Stimuli were generated in Python with the NumPy 1.19.2 and SciPy 1.60 packages. All statistical analyses were carried out in Python with the Pandas 1.2.3, NumPy 1.19, SciPy 1.60, Statsmodels 0.13 and Scikit-Learn 0.24.1 packages. Figures were generated with Matplotlib 3.3.2.

**Code and Data Availability**

All code to reproduce the simulations is available on GitHub under:
https://github.com/summerfieldlab/Flesch_Nagy_etal_HebbCL
Behavioural data from the previously published study is available under**:**
https://github.com/summerfieldlab/flesch_etal_2018

**Stimulus Design**

Stimuli were grayscale images of two-dimensional Gaussian functions with isotropic covariance. We varied the mean of these Gaussian "blobs" in five discrete steps along the x- and y-coordinate, creating a 5x5 grid of possible stimulus locations inside these image patches. The Gaussian blobs were partially overlapping. This gave the network some information about the two-dimensional structure of the stimulus space, which would not have been the case with a conventional one-hot encoding of stimuli.

**Task Design**

We trained feedforward neural networks on a context-dependent decision-making problem, where only a single dimension of the Gaussian blobs (either the x-or y-location) was relevant for each task/context. Each task was defined by a category boundary that divided this space either along the horizontal (first task) or vertical axis (second task). In each task, the network had to learn to "accept" stimuli from one category and "reject" stimuli from the other category.

**Neural Network Architecture**

For all simulations, we used a feed-forward neural network with 25 input units (for the flattened and downscaled grayscale images) and two additional task units, a hidden layer with 100 Rectified Linear Unit (ReLU) non-linearities and a sigmoidal output unit. Weights from the input to the hidden layer were initialised with draws from a zero-mean Gaussian distribution with variance $\sigma^2 = 0.01$. Readout weights were initialised with draws from a zero-mean Gaussian with variance $\sigma^2 = \frac{1}{\sqrt{n\ hidde}}$. All biases were initialised to zero.

**Training Procedures**

All networks were trained on 10000 trials, 5000 per task. In the interleaved curriculum, trials from both tasks were randomly shuffled. In the blocked curriculum, the networks were first trained on all 500 trials from one task, and then on all trials from the other task. Following our previous publication [25], we used a custom loss function which was -1 times the reward associated with "accepting" a Gaussian blob. This was implemented by multiplying the output of the network function (which was in the range 0 to 1 due to the sigmoid) with -R:

$$J(\theta) = -f(x,\theta)R$$

Rewards ranged from -2 to 2 in steps of 1, hence covering all 5 levels of the feature value along the relevant dimension. Hence, the network was encouraged to "accept" rewarding and "reject" non-rewarding stimuli. At the end of the training phase, we evaluated the network's performance on 50 test trials spanning all combinations of task (2), x-position (5) and y-position (5) of the stimuli. For each simulation, we collected 50 independent training runs with randomly initialised neural networks.

**Baseline Model:** The baseline network was trained with vanilla Gradient Descent, applied via Backpropagation to all network weights after each trial:

$$W^{t+1} \leftarrow W^t - \epsilon \nabla_{W^t} J(x_t, \theta)$$

with a learning rate of $\epsilon = 0.2$ for the interleaved and $\epsilon = 0.03$ for the blocked curriculum.

**Sluggishness:** We modelled the "sluggishness" property of the task signals with an Exponentially Moving Average (EMA), which was applied to the task units on each trial. The EMA has the following recursive definition:

$$x_t^{EMA} = \begin{cases} x_1 & for\ t = 1 \\ (1-\alpha)x_t + \alpha x_{t-1}^{EMA} & for\ t > 1 \end{cases}$$

where the hyperparameter $\alpha$ controls the extent to which information from previous trials is carried over to the current trial. To investigate the impact of the sluggishness on task performance, we trained the baseline model (see above) on an interleaved curriculum for a linearly spaced range of 20 $\alpha$ values ranging from 0 to 0.95 and a fixed learning rate of $\epsilon = 0.2$. We collected 50 independent training runs with randomly initialised networks for each of these values.

**Continual Learning with manual gating:** To investigate the impact of non-linear gating on continual task performance, we manually set the weights connecting the task units with each hidden unit to values with opposing signs. More specifically, all "odd" hidden units received a negative bias in the first task and positive bias in the second, whereas all "even" hidden units received the opposite:

$$w_i^h = \begin{cases} [1,-1] & for\ i\ \in \{1,3,5\ ...,n-1\} \\ [-1,1] & for\ i\ \in \{2,4,6,...,n\} \end{cases}$$

We trained the remaining weights of the network with vanilla SGD, just as described for the baseline model above. The learning rate was set to $\epsilon = 0.01$. The network was trained on a blocked curriculum, and we collected 50 independent training runs.

**Continual Learning with Hebbian updates and SGD:** To protect against interference under blocked training, we devised a novel training procedure which consisted of alternating the standard SGD update and a Hebbian learning step. The Hebbian update enabled the network to strengthen associations between the task units and hidden units that carried task-relevant information, while suppressing the output of units with task-irrelevant information. In the following, we motivate this solution from well-known first principles. Hebbian learning strengthens connections between units that are co-activated. Given inputs $x$ and linear hidden units $y$ connected to the inputs via weight matrix $W$ as follows, where $j$ indexes the hidden unit and $i$ the input unit:

$$y_j(x) = \sum_{i=1}^{n} w_i x_i = w_j^T x$$

Hebbian learning performs weight updates proportional to the co-activation of $x$ and $y_j$:

$$\Delta w_{ij} = \eta x_i y_j = \eta x_i \sum_{i=1}^{n} w_i x_i$$

or for the entire vector of weights from inputs to a single hidden unit:

$$\Delta w_j = \eta x y_j = \eta x \sum_{i=1}^{n} w_i x_i$$

The weight updates for standard Hebbian learning are unbounded, which means that weights continue to grow as training progresses. A conventional solution to this problem is to introduce weight decay, leading to the well-known Oja's rule [61]:

$$\Delta w_j = \eta y_j (x - w_j y_j)$$

Oja's rule converges to the first principal component of the dataset, such that $w$ encodes the first eigenvector and $y$ the first eigenvalue of the input covariance matrix [61]. This can be seen by slightly rearranging the terms. In the following, for convenience, we re-derive the analogy between PCA and Oja's rule for the interested reader. First, in the classical formulation of Hebbian learning, we set the learning rate $\eta$ to 1 and introduce an average over multiple trials. Below, $i$ indexes a single input unit for which we compute the correlation with a single hidden unit $y$, and $j$ is the index running over all input units from 1 to $n$:

$$\Delta w_i = \langle y x_i \rangle$$

$$\Delta w_i = \langle \sum_{j}^{n} w_j x_j x_i \rangle = \sum_{j}^{n} w_j \langle x_j x_i \rangle$$

$$\Delta w = \langle x x^T \rangle w = C w$$

With this formulation, the growth of weights $w$ depends solely on the input-input correlation matrix $C$. Now recall that the update equation for Oja's rule is given by

$$\Delta w = y(x - wy) = yx - y^2 w$$

Introducing the average over multiple examples yields

$$\Delta w = \langle yx \rangle - \langle y^2 \rangle w$$

The equilibrium for this equation is reached when the first term on the right is equal to the second, or in other words, when:

$$\langle yx \rangle = \langle y^2 \rangle w$$
$$Cw = \langle y^2 \rangle w$$

From the definition of eigenvalues, it follows that $w$ is an eigenvector of $C$ and $\langle y^2 \rangle = \sigma^2$ its corresponding eigenvalue. Further, the dynamics grow fastest in the direction of the eigenvector with maximal eigenvalue, such that $w$ will converge to the largest principal component of the input data. Applied to the blobs task, this means that weights from the task units to some of the hidden units are positive for one task and negative for the other, while the opposite is true for other units. Together with the supervised learning step, this should allow the network to strengthen positive weights between the active task units and task-relevant hidden units, and negative weights between this task unit and task-irrelevant hidden units. Once the network is exposed to a new task, the opposite mapping should be learned for connections between the second task unit and the hidden layer. We implemented this procedure as follows. For each trial and corresponding input sample $x_t$, we first applied the standard SGD update via backpropagation to all network parameters:

$$W^{t+1} \leftarrow W^t - \epsilon \nabla_{W^t} J(x_t, \theta)$$

This was then followed by a Hebbian update to the weights from the task units to the hidden layer, where y corresponds to the hidden layer activation of the j-th hidden unit prior to the non-linearity and each $w_j$ corresponds to a vector of weights from all task-units to the j-th hidden unit:

$$w_j^{t+1} \leftarrow w_j^t + \eta y_j (x_i - w_j^t y_j)$$

We trained the network on a blocked curriculum as described above, with a learning rate of $\epsilon = 0.0377$ for the SGD and $\eta = 0.00021$ for Hebbian updates with Oja's rule. We collected 50 training runs with independent random initialisations of all network parameters. One might object that mean-centring the task-signal introduces knowledge about the second task during training on the first, as the one-hot inputs [1,0] were converted to [0.5, -0.5]. To overcome this, we used a one-hot signal for the first task and introduced a mean-centred signal for the second task during training. Semantically, this would correspond to first learning how to perform the first task, and then how to do the second task while suppressing information learned about the first.

**Modelling human continual learning:** To model human continual learning, we reduced the number of training trials to 200 per task and combined the sluggishness and Hebbian update procedure outlined above as follows: On each trial, The task signal received by the network was mixed with the signal carried over from previous trials:

$$x_t^{EMA} = \begin{cases} x_1 & for\ t = 1 \\ (1-\alpha)x_t + \alpha x_{t-1}^{EMA} & for\ t > 1 \end{cases}$$

Next, we performed a forward pass through the network and calculated the loss as -1*R:

$$J(\theta) = -\sigma(x,\theta) * R$$

This was then used to perform an SGD update of the network parameters, with a learning rate of $\epsilon = 0.0905$ for the blocked curriculum and $\epsilon = 0.0926$ for the interleaved curriculum:

$$W^{t+1} \leftarrow W^t - \epsilon \nabla_{W^t} J(x_t, \theta)$$

Lastly, the task weights were updated with Oja's rule, with a learning rate of $\eta = 0.0026$ for the blocked and $\eta = 0.000327$ for the interleaved curriculum:

$$w_j^{t+1} \leftarrow w_j^t + \eta y_j(x_i - w_j^t y_j)$$

Notably, in all of these comparisons with human data, we trained the neural networks on the same number of training trials as the human participants from the previous study (200 training trials per task). We performed an extensive hyperparameter search to find learning rates for which the networks would reliably reach ceiling performance (**see also Fig. S2**).
In contrast to the neural network model, Human participants never performed at ceiling on test trials with novel stimuli, not even after extensive training on the tasks. To model this residual cost, we introduced decision noise at test by passing the network's logits through a sigmoid with temperature parameter T that controlled its sensitivity to changes in the input:

$$\sigma(x) = \frac{1}{1 - exp(-\frac{x}{T})}$$

At test, we sampled 10000 choices per input from the trained model by comparing its output to a random uniform variable $X \sim U(0,1)$:

$$y = \begin{cases} 1 & for\ \sigma(x, W) \geq X \\ 0 & for\ \sigma(x, W) < X \end{cases}$$

To fit this model to human choices, we performed a grid search over a range of values for the α and *T* parameters that controlled the amount of sluggishness and decision noise respectively and chose those values that produced outputs which closely resembled the choices made by human participants.

## Quantification and Statistical Analyses

**Test Accuracy:** To compute accuracy during training and test, we evaluated whether the network accepted the rewarding and rejected the non-rewarding trials. Excluding the boundary trials for which the decisions were arbitrary, accuracy was calculated as follows:

$$p(correct) = \frac{1}{n}\sum_i \mathbf{1}_{f(x_i, W) > 0.5 == R_i > 0}$$

**Choice Matrices:** To visualise the choices made by the network, we averaged outputs across trials for each of the 50 unique types of test trials (5 x-positions, 5 y-positions, 2 tasks) and rearranged these outputs into two 5x5 matrices where each entry corresponds to the fraction of "accept" responses for this type of stimulus.

**Task Selectivity:** We performed a regression-based analysis to determine task-selectivity of individual neurons. We regressed their activity against four predictors, coding for the value of relevant and irrelevant feature dimensions of each trial, separately for each task:

$$y_{unit} = \beta_0 + \beta_1 \, relDim_{1st\ task} + \beta_2 \, irrelDim_{1st\ task} + \beta_3 \, relDim_{2nd\ task} + \beta_4 irrelDim_{2nd\ task}$$

Following procedures explained in detail in [15], we defined a unit as being task-selective if its output scaled with the feature value along the relevant – but not irrelevant - dimension of one task, and was zero for the other task. This definition results directly from the rectifying property of ReLUs, which are linear for positive inputs and return zero for negative inputs. It only counts those units as task-selective that have receptive fields aligned with task-relevant information and doesn't consider units that happen to be active in one task, but not the other.

**Hidden layer Representational Similarity Analysis (RSA):** We performed representational similarity analysis (RSA) to investigate the geometry of hidden layer activity patterns of the trained neural networks. First, we collected activity patterns for all 50 conditions (5 x-positions, 5 y-positions, 2 tasks), yielding a 50-x-n_hidden matrix of activity patterns for each individual training run. Next, we created 50x50 representational dissimilarity matrices (RDMs) by computing the pairwise Euclidean distance between all 50 patterns. For visualisation purposes, we then averaged these RDMs across training runs (separately for the blocked and interleaved curriculum) and projected them down into 3 dimensions using classical Multi-Dimensional Scaling (MDS). As MDS is rotation-invariant, we manually rotated the resulting projection so that axes of the projection were aligned with the figure axes, which made it easier to compare the geometry across conditions (and models). To get quantitative insights into the geometry of these patterns, we regressed these RDMs against a set of model RDMs that encoded (a) grid-like, (b) orthogonal or (c) diagonal patterns at the level of individual runs. Let the vectors for x- and y-position be $x = [-2,-1,0,1,2]^T$ and $y = [-2,-1,0,1,2]^T$. Let the task vector be defined as $t = [0,1]^T$. Let the matrix of all possible ordered tuples of context, x- and y-position be:

$$X^{50x3} = \{(ti, xi, yi) : ti \in t;\ xi \in x\ and\ yi \in y\ \}$$

The grid-like RDM was constructed by computing the pairwise Euclidean distances between all rows of X.

The orthogonal model RDM was obtained by projecting stimuli into task-relevant axes, so that only the x-position was encoded for the first task, and only the y-position for the second task, leaving a representation where two orthogonal one-dimensional manifolds were separated along a third axis that encoded the task. Let $X_A$ be the submatrix for the first task and $X_B$ the submatrix for the second task:

$$X_{grid} = \begin{bmatrix} X_A \\ X_B \end{bmatrix}$$

Let $Y_A$ be the projection matrix for the first task and $Y_B$ the projection matrix for the second task:

$$Y_A = \begin{bmatrix} 1 & 0 & 0 \\ 0 & 0 & 0 \\ 0 & 0 & 1 \end{bmatrix}$$

$$Y_B = \begin{bmatrix} 1 & 0 & 0 \\ 0 & 1 & 0 \\ 0 & 0 & 0 \end{bmatrix}$$

Then, the orthogonal model corresponded to stacking $X_A Y_A$ and $X_B Y_B$:

$$X_{orth} = \begin{bmatrix} X_A Y_A \\ X_B Y_B \end{bmatrix}$$

The diagonal model corresponded to a neural representation that only differentiated between stimuli along the diagonal from low x- and y-values to high x- and y-values. This assumed that participants learned a single boundary for both tasks and optimised for a strategy that led to 70% correct in both tasks. We constructed this diagonal model RDM with the projection $XP^T$ where:

$$P = \begin{bmatrix} 1 & 0 \\ 0 & cos(45) \\ 0 & sin(45) \end{bmatrix}$$

$$X_{diag} = X_{grid} P^T$$

To estimate the extent to which each of these models explained the geometry of representations in the hidden layers of our neural networks, we performed a multiple linear regression at the level of individual runs, in which we regressed the hidden layer RDM against the set of model RDMs, after z-scoring and vectorising the lower-triangular form of each RDM:

$$RDM_{hidden} = \beta_0 + \beta_1 RDM_{grid} + \beta_2 RDM_{orth} + \beta_3 RDM_{diag}$$

For statistical inference at the group-level, we performed t-tests against zero on each set of regression coefficients.

**Comparison with Human behavioural data.** We followed procedures described in [25] for our re-analysis of the behavioural data. In the original study, there were four groups that differed in the amount of "blockiness" during training, ranging from a fully blocked curriculum where participants were trained on one task and then the other, to a fully interleaved curriculum in which trials were randomly interspersed. In our re-analysis, we focus on the two extremes, called the "blocked 200" group and "interleaved" group in the original publication. As the calculation of the sigmoidal fits, model-based RSA and fits of the psychophysical model were identical to those described in the original paper, we're providing an abbreviated version of the methods below.

**Sigmoid fits:** To estimate sensitivity of choices made by the networks/human participants to the relevant and irrelevant feature dimensions, we fit sigmoidal curves at the level of individual runs/participants. First, responses were averaged across test trials and tasks within each of the five bins along a given dimension. Next, we fit a sigmoidal curve of the following form to the data, using the curve_fit function of the SciPy package:

$$\sigma(x) = \frac{L + (1 - 2L)}{1 - exp(-k(x - x0))}$$

where L controlled the proportion of nonspecific errors (lapses), k the slope and x0 the offset of the sigmoid. Statistical inference was performed on the group-level distributions of the individually estimated parameters.

**Factorised/linear model:** To calculate the extent to which the neural networks/human participants learned a factorised solution, comprised of one accurate category boundary per task, or a linear solution, where the same boundary was applied to both tasks, we performed a model-based representational similarity analysis on the network outputs / behaviour. First, we created choice matrices (see above) for each network run / at single subject level. We then constructed two model choice matrices, the factorised and the linear model. In the factorised model, all entries corresponding to rewarding trials were set to 1, and entries corresponding to non-rewarding trials were set to zero. Category-boundary trials were set to 0.5. In the linear model, we assumed a diagonal category boundary distinguishing between trials that were rewarding/non-rewarding irrespective of context and set the corresponding entries in the two matrices to 1, 0.5 and 0 respectively. We then concatenated the flattened choice matrices for the first and second task and constructed RDMs from the resulting vectors using the squareform and pdist functions from the SciPy package. The empirical RDMs, constructed from the network output / human behaviour were then regressed against the two model RDMs at the level of single runs / subjects.

**Psychophysical model:** To decompose errors made by the neural networks / human participants into different sources, we fit a psychophysical model with five free parameters to individual runs / participants. The model had parameters for the angles of the decision boundaries in the two-dimensional stimulus space, as well as the slope, offset and lapse-rate of a sigmoidal transducer. The model projected the 2D stimulus space onto an axis perpendicular to the decision boundary and fed the projected values through a sigmoid to generate choice probabilities. Let $X_a$ and $X_b$ be the 25x2 matrices of coordinates for the stimuli of the first and second task, where each row corresponds to the x- and y-location of the peak of a Gaussian "blob". The first two free parameters $\theta_a$ and $\theta_b$ determined the angle of the line onto which these stimuli were projected:

$$X^{proj} = \begin{bmatrix} X_a[cos(\theta_a), sin(\theta_a)]^T \\ X_b[cos(\theta_b), sin(\theta_b)]^T \end{bmatrix}$$

Next, the projected values were passed through a sigmoidal transducer with free parameters for the lapse rate L, the slope k and the offset x0:

$$\hat{y}(X^{proj}) = \frac{L + (1 - 2L)}{1 - exp(-k(X^{proj} - x0))}$$

We fit this model to empirical data by minimising the following loss function that quantified the mismatch between the model's output and the choices made by the network / human participant:

$$J(y, \hat{y}; \theta, L, k, x0) = -\sum_i log(1 - |y_i - \hat{y}_i(\theta, L, k, x0)|)$$

Minimisation was performed with the L-BFGS algorithm as implemented in the minimise function of the SciPy package, with constraints set on the range of parameter values so that angle theta in [0,359], slope in [0,20], offset in [-1,1] and lapse in [0, 0.5].

## ACKNOWLEDGEMENTS

This work was supported by generous funding from the European Research Council (ERC Consolidator award 725937) and Special Grant Agreement No. 945539 (Human Brain Project SGA) to C.S., a Sir Henry Dale Fellowship to A.S. from the Wellcome Trust and Royal Society (grant number 216386/Z/19/Z), and a Medical Science Graduate School Studentship and funding from the Covid Scholarship Extension Fund to T.F. (Medical Research Council and Department of Experimental Psychology). A.S. is a CIFAR Azrieli Global Scholar in the Learning in Machines & Brains program. D.G.N. received support from the National Research, Development and Innovation Fund of Hungary (No. K125343, awarded to Gergo Orban, https://nkfih.gov.hu/about-the-office).

## DECLARATION OF INTERESTS

The authors declare no competing interests.

# REFERENCES

1. Parisi GI, Kemker R, Part JL, Kanan C, Wermter S. Continual lifelong learning with neural networks: A review. Neural Networks. 2019;113: 54–71. doi:10.1016/j.neunet.2019.01.012

2. Hadsell R, Rao D, Rusu AA, Pascanu R. Embracing Change: Continual Learning in Deep Neural Networks. Trends in Cognitive Sciences. 2020;24: 1028–1040. doi:10.1016/j.tics.2020.09.004

3. Musslick S, Cohen JD. Rationalizing constraints on the capacity for cognitive control. Trends in Cognitive Sciences. 2021;25: 757–775. doi:10.1016/j.tics.2021.06.001

4. Franklin NT, Frank MJ. Generalizing to generalize: Humans flexibly switch between compositional and conjunctive structures during reinforcement learning. PLOS Computational Biology. 2020;16: e1007720. doi:10.1371/journal.pcbi.1007720

5. Wulf G, Shea CH. Principles derived from the study of simple skills do not generalize to complex skill learning. Psychonomic Bulletin & Review. 2002;9: 185–211. doi:10.3758/BF03196276

6. Carvalho PF, Goldstone RL. Putting category learning in order: Category structure and temporal arrangement affect the benefit of interleaved over blocked study. Mem Cogn. 2014;42: 481–495. doi:10.3758/s13421-013-0371-0

7. Carvalho PF, Goldstone RL. What you learn is more than what you see: what can sequencing effects tell us about inductive category learning? Front Psychol. 2015;6: 505. doi:10.3389/fpsyg.2015.00505

8. Richards BA, Lillicrap TP, Beaudoin P, Bengio Y, Bogacz R, Christensen A, et al. A deep learning framework for neuroscience. Nature Neuroscience. 2019;22: 1761–1770. doi:10.1038/s41593-019-0520-2

9. Saxe A, Nelli S, Summerfield C. If deep learning is the answer, what is the question? Nature Reviews Neuroscience. 2021;22: 55–67. doi:10.1038/s41583-020-00395-8

10. Yamins DLK, Hong H, Cadieu CF, Solomon EA, Seibert D, DiCarlo JJ. Performance-optimized hierarchical models predict neural responses in higher visual cortex. PNAS. 2014;111: 8619–8624. doi:10.1073/pnas.1403112111

11. Khaligh-Razavi S-M, Kriegeskorte N. Deep Supervised, but Not Unsupervised, Models May Explain IT Cortical Representation. PLOS Computational Biology. 2014;10: e1003915. doi:10.1371/journal.pcbi.1003915

12. Güçlü U, Gerven MAJ van. Deep Neural Networks Reveal a Gradient in the Complexity of Neural Representations across the Ventral Stream. J Neurosci. 2015;35: 10005–10014. doi:10.1523/JNEUROSCI.5023-14.2015

13. Lindsay GW. Convolutional Neural Networks as a Model of the Visual System: Past, Present, and Future. Journal of Cognitive Neuroscience. 2021;33: 2017–2031. doi:10.1162/jocn_a_01544

14. Zhuang C, Yan S, Nayebi A, Schrimpf M, Frank MC, DiCarlo JJ, et al. Unsupervised neural network models of the ventral visual stream. PNAS. 2021;118. doi:10.1073/pnas.2014196118

15. Flesch T, Juechems K, Dumbalska T, Saxe A, Summerfield C. Orthogonal representations for robust context-dependent task performance in brains and neural networks. Neuron. 2022;0.

doi:10.1016/j.neuron.2022.01.005

16. Yang GR, Joglekar MR, Song HF, Newsome WT, Wang X-J. Task representations in neural networks trained to perform many cognitive tasks. Nat Neurosci. 2019;22: 297–306. doi:10.1038/s41593-018-0310-2

17. Ito T, Murray JD. Multi-task representations in human cortex transform along a sensory-to-motor hierarchy. bioRxiv. 2021; 2021.11.29.470432.

18. Badre D, Bhandari A, Keglovits H, Kikumoto A. The dimensionality of neural representations for control. Current Opinion in Behavioral Sciences. 2021;38: 20–28. doi:10.1016/j.cobeha.2020.07.002

19. Jagadeesh AV, Gardner JL. Texture-like representation of objects in human visual cortex. bioRxiv. 2022; 2022.01.04.474849.

20. French RM. Catastrophic forgetting in connectionist networks. Trends in Cognitive Sciences. 1999;3: 128–135. doi:10.1016/S1364-6613(99)01294-2

21. Lee S, Goldt S, Saxe A. Continual Learning in the Teacher-Student Setup: Impact of Task Similarity. Proceedings of the 38th International Conference on Machine Learning. PMLR; 2021. pp. 6109–6119. Available: https://proceedings.mlr.press/v139/lee21e.html

22. Ehret B, Henning C, Cervera MR, Meulemans A, von Oswald J, Grewe BF. Continual Learning in Recurrent Neural Networks. arXiv; 2021. doi:10.48550/arXiv.2006.12109

23. Mnih V, Kavukcuoglu K, Silver D, Rusu AA, Veness J, Bellemare MG, et al. Human-level control through deep reinforcement learning. Nature. 2015;518: 529–533. doi:10.1038/nature14236

24. Kirkpatrick J, Pascanu R, Rabinowitz N, Veness J, Desjardins G, Rusu AA, et al. Overcoming catastrophic forgetting in neural networks. PNAS. 2017;114: 3521–3526. doi:10.1073/pnas.1611835114

25. Flesch T, Balaguer J, Dekker R, Nili H, Summerfield C. Comparing continual task learning in minds and machines. PNAS. 2018;115: E10313–E10322. doi:10.1073/pnas.1800755115

26. Hadsell R, Rao D, Rusu AA, Pascanu R. Embracing Change: Continual Learning in Deep Neural Networks. Trends in Cognitive Sciences. 2020;24: 1028–1040. doi:10.1016/j.tics.2020.09.004

27. Zhang Y, Yang Q. An overview of multi-task learning. National Science Review. 2017;5: 30–43. doi:10.1093/nsr/nwx105

28. Zenke F, Poole B, Ganguli S. Continual Learning Through Synaptic Intelligence. arXiv:170304200 [cs, q-bio, stat]. 2017 [cited 21 Jan 2022]. Available: http://arxiv.org/abs/1703.04200

29. Rusu AA, Rabinowitz NC, Desjardins G, Soyer H, Kirkpatrick J, Kavukcuoglu K, et al. Progressive Neural Networks. arXiv:160604671 [cs]. 2016 [cited 21 Jan 2022]. Available: http://arxiv.org/abs/1606.04671

30. Shin H, Lee JK, Kim J, Kim J. Continual Learning with Deep Generative Replay. Advances in Neural Information Processing Systems. Curran Associates, Inc.; 2017. Available: https://proceedings.neurips.cc/paper/2017/hash/0efbe98067c6c73dba1250d2beaa81f9-Abstract.html

31. Farajtabar M, Azizan N, Mott A, Li A. Orthogonal Gradient Descent for Continual Learning. arXiv:191007104 [cs, stat]. 2019 [cited 21 Jan 2022]. Available: http://arxiv.org/abs/1910.07104

32. Zeng G, Chen Y, Cui B, Yu S. Continual learning of context-dependent processing in neural networks. Nat Mach Intell. 2019;1: 364–372. doi:10.1038/s42256-019-0080-x

33. Chaudhry A, Khan N, Dokania PK, Torr PHS. Continual Learning in Low-rank Orthogonal Subspaces. arXiv:201011635 [cs]. 2020 [cited 21 Jan 2022]. Available: http://arxiv.org/abs/2010.11635

34. Duncker L, Driscoll L, Shenoy KV, Sahani M, Sussillo D. Organizing recurrent network dynamics by task-computation to enable continual learning. Advances in Neural Information Processing Systems. Curran Associates, Inc.; 2020. pp. 14387–14397. Available: https://proceedings.neurips.cc/paper/2020/hash/a576eafbce762079f7d1f77fca1c5cc2-Abstract.html

35. Liu P, Qiu X, Huang X. Recurrent Neural Network for Text Classification with Multi-Task Learning. arXiv; 2016. doi:10.48550/arXiv.1605.05101

36. van de Ven GM, Siegelmann HT, Tolias AS. Brain-inspired replay for continual learning with artificial neural networks. Nat Commun. 2020;11: 4069. doi:10.1038/s41467-020-17866-2

37. McClelland JL, McNaughton BL, O'Reilly RC. Why there are complementary learning systems in the hippocampus and neocortex: insights from the successes and failures of connectionist models of learning and memory. Psychol Rev. 1995;102: 419–457. doi:10.1037/0033-295X.102.3.419

38. Masse NY, Grant GD, Freedman DJ. Alleviating catastrophic forgetting using context-dependent gating and synaptic stabilization. PNAS. 2018;115: E10467–E10475. doi:10.1073/pnas.1803839115

39. Kaplanis C, Shanahan M, Clopath C. Continual Reinforcement Learning with Complex Synapses. arXiv:180207239 [cs]. 2018 [cited 21 Jan 2022]. Available: http://arxiv.org/abs/1802.07239

40. Benna MK, Fusi S. Computational principles of synaptic memory consolidation. Nat Neurosci. 2016;19: 1697–1706. doi:10.1038/nn.4401

41. Libby A, Buschman TJ. Rotational dynamics reduce interference between sensory and memory representations. Nature Neuroscience. 2021; 1–12. doi:10.1038/s41593-021-00821-9

42. Panichello MF, Buschman TJ. Shared mechanisms underlie the control of working memory and attention. Nature. 2021;592: 601–605. doi:10.1038/s41586-021-03390-w

43. Miller EK, Cohen JD. An integrative theory of prefrontal cortex function. Annu Rev Neurosci. 2001;24: 167–202. doi:10.1146/annurev.neuro.24.1.167

44. Rougier NP, Noelle DC, Braver TS, Cohen JD, O'Reilly RC. Prefrontal cortex and flexible cognitive control: Rules without symbols. PNAS. 2005;102: 7338–7343. doi:10.1073/pnas.0502455102

45. Rikhye RV, Gilra A, Halassa MM. Thalamic regulation of switching between cortical representations enables cognitive flexibility. Nat Neurosci. 2018;21: 1753–1763. doi:10.1038/s41593-018-0269-z

46. Johnston K, Levin HM, Koval MJ, Everling S. Top-down control-signal dynamics in anterior cingulate and prefrontal cortex neurons following task switching. Neuron. 2007;53: 453–462. doi:10.1016/j.neuron.2006.12.023

47. Mansouri FA, Matsumoto K, Tanaka K. Prefrontal Cell Activities Related to Monkeys' Success and Failure in Adapting to Rule Changes in a Wisconsin Card Sorting Test Analog. J Neurosci. 2006;26: 2745–2756. doi:10.1523/JNEUROSCI.5238-05.2006

48. Buchsbaum BR, Greer S, Chang W-L, Berman KF. Meta-analysis of neuroimaging studies of the Wisconsin card-sorting task and component processes. Hum Brain Mapp. 2005;25: 35–45. doi:10.1002/hbm.20128

49. Cohen JD, Dunbar K, McClelland JL. On the control of automatic processes: A parallel distributed processing account of the Stroop effect. Psychological Review. 1990;97: 332–361. doi:10.1037/0033-295X.97.3.332

50. Gisiger T, Boukadoum M. Mechanisms Gating the Flow of Information in the Cortex: What They Might Look Like and What Their Uses may be. Front Comput Neurosci. 2011;5: 1. doi:10.3389/fncom.2011.00001

51. Tsuda B, Tye KM, Siegelmann HT, Sejnowski TJ. A modeling framework for adaptive lifelong learning with transfer and savings through gating in the prefrontal cortex. PNAS. 2020;117: 29872–29882. doi:10.1073/pnas.2009591117

52. Serrà J, Surís D, Miron M, Karatzoglou A. Overcoming catastrophic forgetting with hard attention to the task. arXiv; 2018. doi:10.48550/arXiv.1801.01423

53. Verbeke P, Verguts T. Using top-down modulation to optimally balance shared versus separated task representations. Neural Networks. 2022;146: 256–271. doi:10.1016/j.neunet.2021.11.030

54. Russin J, Zolfaghar M, Park SA, Boorman E, O'Reilly RC. A Neural Network Model of Continual Learning with Cognitive Control. arXiv:220204773 [cs, q-bio]. 2022 [cited 3 Mar 2022]. Available: http://arxiv.org/abs/2202.04773

55. Soetens E, Boer LC, Hueting JE. Expectancy or automatic facilitation? Separating sequential effects in two-choice reaction time. Journal of Experimental Psychology: Human Perception and Performance. 1985;11: 598–616. doi:10.1037/0096-1523.11.5.598

56. Cho RY, Nystrom LE, Brown ET, Jones AD, Braver TS, Holmes PJ, et al. Mechanisms underlying dependencies of performance on stimulus history in a two-alternative forced-choice task. Cognitive, Affective, & Behavioral Neuroscience. 2002;2: 283–299. doi:10.3758/CABN.2.4.283

57. Yu AJ, Cohen JD. Sequential effects: Superstition or rational behavior? Advances in Neural Information Processing Systems. Curran Associates, Inc.; 2008. Available: https://proceedings.neurips.cc/paper/2008/hash/5f2c22cb4a5380af7ca75622a6426917-Abstract.html

58. Flesch T, Nagy D, Saxe A, Summerfield C. Modelling continual learning in humans with Hebbian context gating. Cosyne Abstracts. 2021.

59. Mante V, Sussillo D, Shenoy KV, Newsome WT. Context-dependent computation by recurrent dynamics in prefrontal cortex. Nature. 2013;503: 78–84. doi:10.1038/nature12742

60. Monsell S. Task switching. Trends in Cognitive Sciences. 2003;7: 134–140. doi:10.1016/S1364-6613(03)00028-7

61. Oja E. Simplified neuron model as a principal component analyzer. J Math Biology. 1982;15: 267–273. doi:10.1007/BF00275687

62. Oja E, Karhunen J. On stochastic approximation of the eigenvectors and eigenvalues of the expectation of a random matrix. Journal of Mathematical Analysis and Applications. 1985;106: 69–84. doi:10.1016/0022-247X(85)90131-3

63. O'Reilly RC, Frank MJ. Making working memory work: a computational model of learning in the prefrontal cortex and basal ganglia. Neural Comput. 2006;18: 283–328. doi:10.1162/089976606775093909

64. Herd SA, O'Reilly RC, Hazy TE, Chatham CH, Brant AM, Friedman NP. A neural network model of individual differences in task switching abilities. Neuropsychologia. 2014;62: 375–389. doi:10.1016/j.neuropsychologia.2014.04.014

65. Xie Y, Hu P, Li J, Chen J, Song W, Wang X-J, et al. Geometry of sequence working memory in macaque prefrontal cortex. Science. 2022;375: 632–639. doi:10.1126/science.abm0204

66. Postle BR. Delay-period activity in prefrontal cortex: one function is sensory gating. J Cogn Neurosci. 2005;17: 1679–1690. doi:10.1162/089892905774589208

67. Vander Weele CM, Siciliano CA, Matthews GA, Namburi P, Izadmehr EM, Espinel IC, et al. Dopamine enhances signal-to-noise ratio in cortical-brainstem encoding of aversive stimuli. Nature. 2018;563: 397–401. doi:10.1038/s41586-018-0682-1

68. Jensen O, Mazaheri A. Shaping Functional Architecture by Oscillatory Alpha Activity: Gating by Inhibition. Frontiers in Human Neuroscience. 2010;4. Available: https://www.frontiersin.org/articles/10.3389/fnhum.2010.00186

69. Servan-Schreiber D, Printz H, Cohen JD. A Network Model of Catecholamine Effects: Gain, Signal-to-Noise Ratio, and Behavior. Science. 1990;249: 892–895. doi:10.1126/science.2392679

70. Iyer A, Grewal K, Velu A, Souza LO, Forest J, Ahmad S. Avoiding Catastrophe: Active Dendrites Enable Multi-Task Learning in Dynamic Environments. Front Neurorobot. 2022;16: 846219. doi:10.3389/fnbot.2022.846219

71. Grewal K, Forest J, Cohen BP, Ahmad S. Going Beyond the Point Neuron: Active Dendrites and Sparse Representations for Continual Learning. bioRxiv; 2021. p. 2021.10.25.465651. doi:10.1101/2021.10.25.465651

72. Rohrer D, Dedrick RF, Stershic S. Interleaved practice improves mathematics learning. Journal of Educational Psychology. 2015;107: 900–908. doi:10.1037/edu0000001

73. Samani J, Pan SC. Interleaved practice enhances memory and problem-solving ability in undergraduate physics. npj Sci Learn. 2021;6: 1–11. doi:10.1038/s41539-021-00110-x

74. Ramasesh VV, Dyer E, Raghu M. Anatomy of Catastrophic Forgetting: Hidden Representations and Task Semantics. 2021. Available: https://openreview.net/forum?id=LhY8QdUGSuw

75. Musslick S, Saxe A, Hoskin AN, Reichman D, Cohen JD. On the Rational Boundedness of

Cognitive Control: Shared Versus Separated Representations. PsyArXiv. 2020 [cited 17 Mar 2022]. Available: https://psyarxiv.com/jkhdf/

# Supporting Information

**Supplementary Methods**

**Neural network simulations with trees stimuli**
To verify that our approach could be extended to slightly more complex input datasets and architectures, we repeated the experiments with MLPs with two hidden layers that were trained on a down-sampled version of the fractal tree images from the original paper.

**Stimulus Design**
Stimuli were images of fractal trees that varied in five discrete steps in terms of their density of branches ("branchiness") and leaves ("leafiness") and were pasted onto a grey background. We took the dataset of 50000 training and 10000 test images that was used in Flesch et al., 2018 and down sampled each image to 24x24x3 pixels. Pixel values were encoded as floats in the range from 0 to 1.

**Neural Network Architecture**
The neural network was similar to the ones described in the main text. The input layer consisted of 24*24*3=1728 units that received the flattened RGB images, and two additional task units which received a one-hot encoded context signal. Inputs were passed through two hidden layers with 100 ReLU non-linearities each. The output was a single node with sigmoid non-linearity.

**Training Procedure**
All training procedures were similar to those described in the main text. We used the same weight initialisation and performed a hyperparameter search over a range of values for the SGD and Hebb-update learning rates, as well as a "context offset" parameter that was multiplied with the one-hot encoded context signal. The purpose of this scaling was to get a high enough activation from the context units, relative to the input units. The baseline network was trained with a learning rate of $\epsilon = 0.001854$ and context offset of of $c = 1$ for interleaved and a learning rate of $\epsilon = 0.001968$ and context offset of $c = 4$ for the blocked curriculum. The Hebbian network was trained with earning rates $\epsilon = 0.001968$ and $\eta = 0.000849$ and context offset of $c = 4$. For comparisons with human data, we trained the sluggish Hebbian network with sluggishness values ranging from $\alpha = 0.05$ to $\alpha = 1.00$ in 30 steps and learning rates $\epsilon = 0.001968$ and $\eta = 0.000849$ and chose the sluggishness value that minimised the difference between choices made by humans and the neural network outputs. For each network, we collected 50 training runs with independent random weight initialisations. To stabilise training of the Hebbian network, we had to restrict Hebbian weight updates to the connections from the two context units to the hidden layer.

All networks were trained for 100 episodes (50 training trials per episode, spanning all 5x5x2 combinations of branchiness, leafiness and tasks, but with randomly drawn specific exemplars) and evaluated on all 10000 test stimuli to assess generalisation performance. All reported results (except for learning curves) are based on those test trials.

# Supplementary Figures

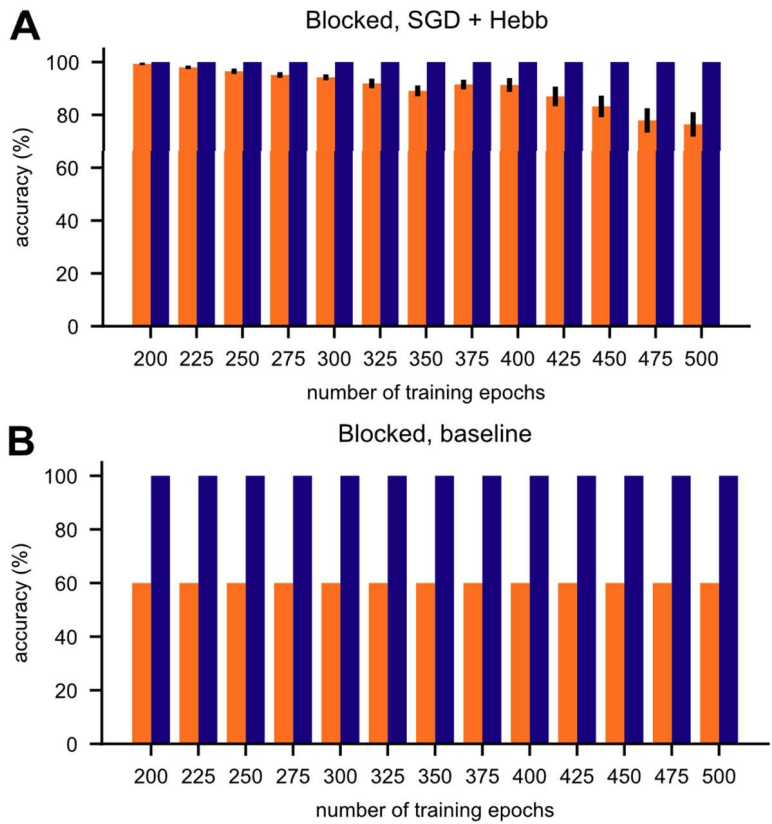

**Figure S1: Impact of block length on forgetting.** (**A**) Impact of block length on forgetting of the first task in network trained with SGD and Hebbian updates on blocked curriculum. Hyperparameters were optimised for a block length of 200 epochs and kept the same for all other training lengths. Even with more than twice as many training epochs, the network was still able to perform the first task well. (**B**) Same as (A) but for network trained without Hebbian updates.

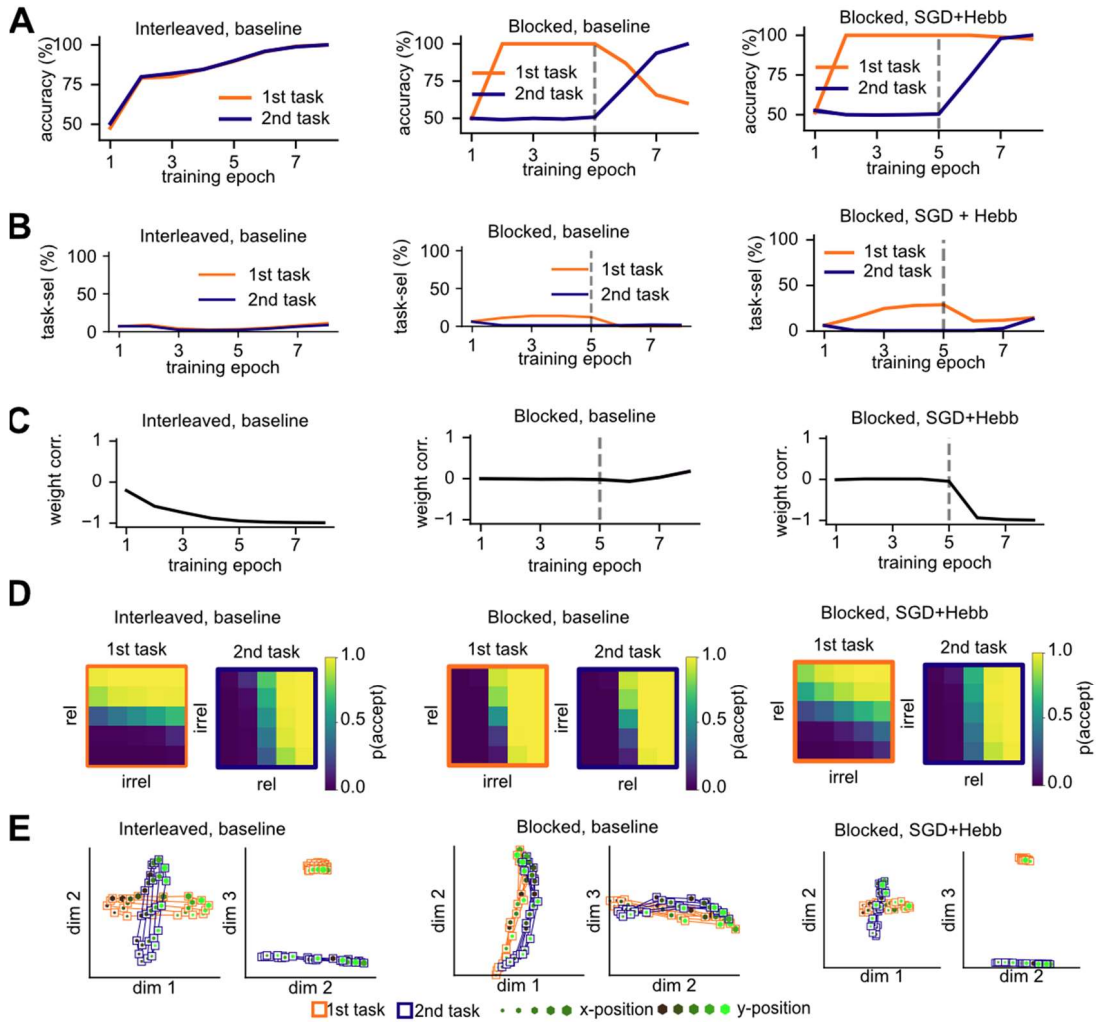

**Figure S2: Replication of key findings with network trained only on 8 episodes.** We repeated the main simulations with the two baseline nets, which were only trained with vanilla SGD, either on interleaved or blocked data, and the network trained both with SGD + Hebbian updates on a blocked curriculum. Hyperparameters were optimised for each net separately. The results show that key findings can be replicated, even if the networks receive as few trials as human participants in the original study. (**A**) Training accuracy for the vanilla net, trained on interleaved or blocked trials, and the Hebbian net, trained on blocked trials. The vanilla net converges on interleaved data but suffers from catastrophic interference under blocked training. In contrast, the Hebbian network's performance on the first task remains at ceiling. (**B**) Fraction of units that became purely task selective. While overall, the fractions were much lower than in the networks trained on 200 episodes, more task-selective units were found in the vanilla network trained on interleaved data and the Hebbian network trained on blocked data, compared to the vanilla network that received a blocked training curriculum. (**C**) Correlation between context weights. Interleaved training with a vanilla network and blocked training with the Hebbian intervention both induced anti-correlated context weights. The vanilla network trained on blocked data failed to utilise the context signal. (**D**) Network responses. The vanilla network trained on blocked data treated the first task as if it was the second. The other two networks learned accurate estimates of the category boundaries. (**E**) MDS on the hidden layer activity patterns. The vanilla network trained on blocked data filtered out the dimension that was irrelevant for the second task but applied the same strategy to the first task. In contrast, the vanilla network trained on interleaved data and the Hebbian network trained on blocked data formed orthogonal representations, also consistent with our previous reports.

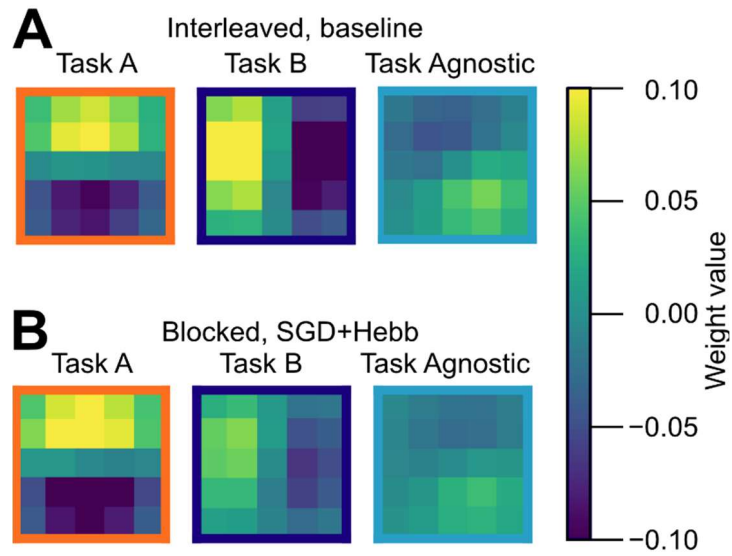

**Figure S3: Weights from input units to task-selective and -agnostic units in the hidden layer.** (**A**) Learned weights for vanilla network, trained on interleaved data. Each heatmap shows averages of the weights from the input layer to hidden units that are selective to either the first or second task, or task-agnostic, reshaped from a 25x1 vector to a 5x5 matrix to resemble the dimensionality of the input images. The plots indicate that task-selective units are associated with weights that select for the task-relevant dimensions (position along the x- and y- axis respectively), while the task-agnostic units code for stimuli that have the same response across tasks (=congruent trials). (**B**) Same as (A) but for network trained additionally with Hebbian updates on a blocked curriculum. A very similar structure was observed.

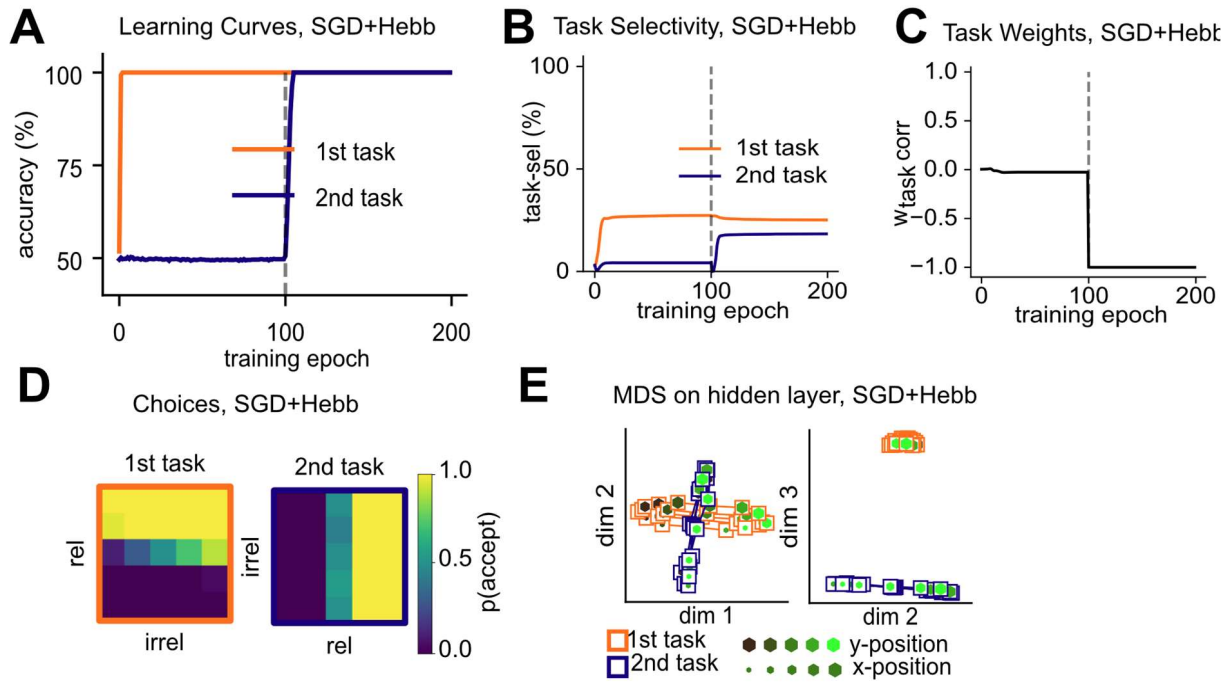

**Figure S4. Continual learning with Hebbian step applied exclusively to task units.** Our approach solves two problems, how to discover the context signal and how to use it to gate out irrelevant dimensions. The former requires that the context/task signal is the largest principal component in the dataset. We note, however, that the mechanism still protects against catastrophic forgetting if the Hebbian updates are only applied to task units (instead of all inputs) and it is assumed that the task signal has already been identified among the inputs. This is shown here. (A) Learning curves for network trained with SGD and Hebbian updates on blocked curriculum. (B) Emergence of task-selective units in the hidden layer. (C) Correlation between weights from task-units in the input layer to the hidden layer. (D) Choices made by the network at the end of training. (E) Projections of learned hidden layer representations into three dimensions.

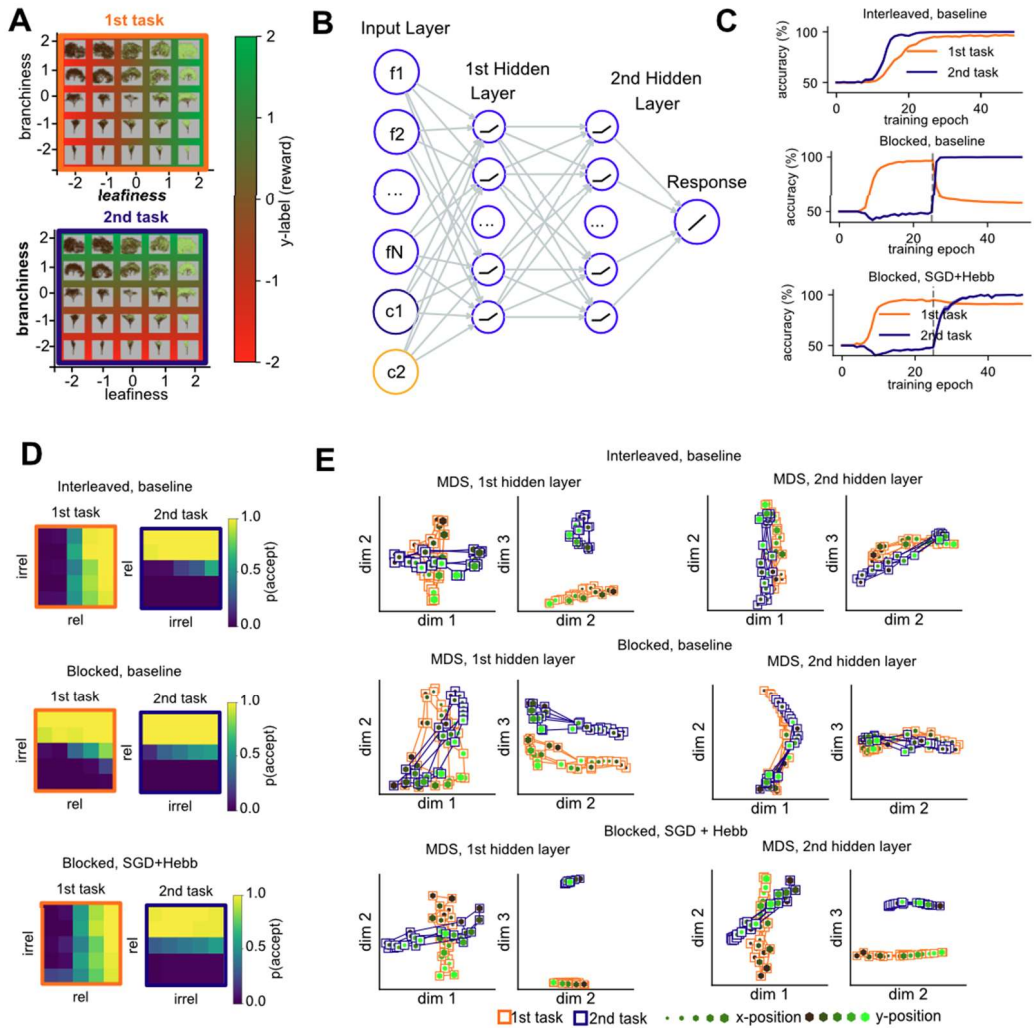

**Figure S5: Replication with tree images and deeper MLP.** (**A**) Stimuli and task rules. Instead of Gaussian blobs, we used RGB images of fractal tree images from the original study. Trees varied in five discrete steps in terms of their density of branches ("branchiness") and leaves ("leafiness"). Only one of the two dimensions was relevant in each context/task. (**B**) Network architecture. Once again, we used a feed-forward neural network, but this time with two hidden layers with ReLU non-linearities. Inputs were flattened and normalised RGB images of trees, together with a one-hot encoded task signal that indicated whether the network was doing the first or the second task. (**C**) Learning curves for vanilla network trained just with SGD on either interleaved (top) or blocked (middle) data, and Hebbian network that was trained on blocked data with SGD and Hebbian updates (bottom). Learning curves were similar to those observed with the simpler network and Gaussian blobs. (**D**) Outputs of the three networks. The vanilla network trained on blocked data treated the first task as if it was the second. The other two networks learned accurate category boundary estimates. (**E**) MDS applied to patterns in both hidden layers of the three networks. The baseline network, trained on interleaved data formed orthogonal representations in the first hidden layer (1st row, left) and parallel representations in its second layer (1st row, right). These parallel representaitons were obtained by rotating one of the task manifolds from the previous layer by 90 degrees, to bring both into the frame of reference of the response, so that leafiness of the first task was mapped onto the same axis as branchiness of the second task. The network trained on blocked data, in contrast, just represented branchiness, which was relevant for the second task, and did not distinguish between contexts (2nd row). The network trained with Hebbian updates on blocked data (3rd row), in contrast, learned similar representations as the baseline network trained on interleaved data.

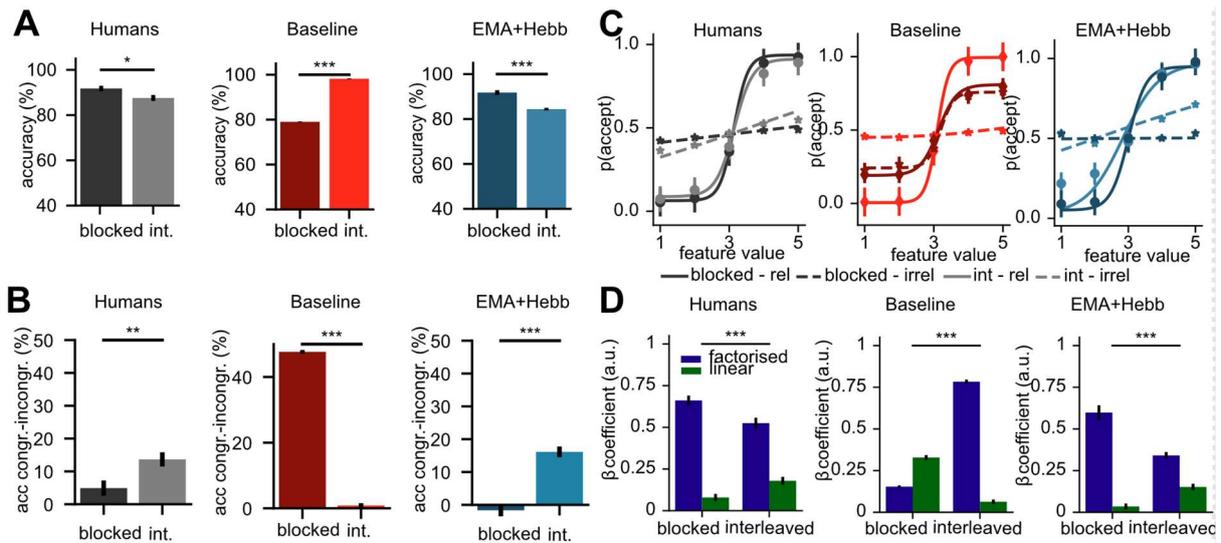

**Figure S6: Comparison of human behavioural results from Flesch et al., 2018 to MLPs trained on fractal tree images.** Same analysis as shown in Fig. 5 (main text), but for MLPs that were trained on the same fractal tree images as human participants. (A) Test phase accuracy. Both humans and the sluggish network trained with additional Hebbian updates (EMA+Hebb) perform better under blocked compared to interleaved training (humans: $T(93) = 2.32$, $p = 0.022$; EMA+Hebb: $T(98) = 3.81$, $p = 0.0005$). The baseline network, only trained with SGD ("Baseline") suffered from catastrophic interference (accuracy blocked < interleaved: $T(98) = 73.94$, $p < 0.0001$). **(B)** Congruency effect. Just like human participants, under interleaved training, our model performs worse on incongruent compared to congruent trials (humans: $T(93) = 2.73$, $p = 0.0075$, EMA+Hebb: $T(98) = 7.03$, $p < 0.0001$). **(C)** Sigmoidal fits of choices made by human participants and the two types of MLPs. The EMA+Hebb network makes choices similar to human participants, with more intrusion from irrelevant dimensions under interleaved compared to blocked training (humans: $T(93) = 28.25$, $p < 0.0001$, EMA+Hebb: $T(98) = 5.13$, $p < 0.0001$). The baseline network doesn't capture human choice patterns. **(D)** Fits of the factorised and linear model to human and network choices. For humans and the EMA+Hebb model, the factorised model fits better under blocked than interleaved training (humans: $T(93) = 3.07$, $p = 0.0028$, EMA+Hebb: $T(98) = 5.14$, $p < 0.0001$), while the opposite was true for the linear model (humans: $T(93) = -3.12$, $p = 0.0024$, EMA+Hebb: $T(98) = -4.39$, $p = 0.0001$). Error patterns of the baseline model show the opposite pattern (Factorised model blocked < interleaved: $T(98) = -40.86$, $p < 0.0001$, linear model blocked > interleaved: $T(98) = 13.06$, $p < 0.0001$).

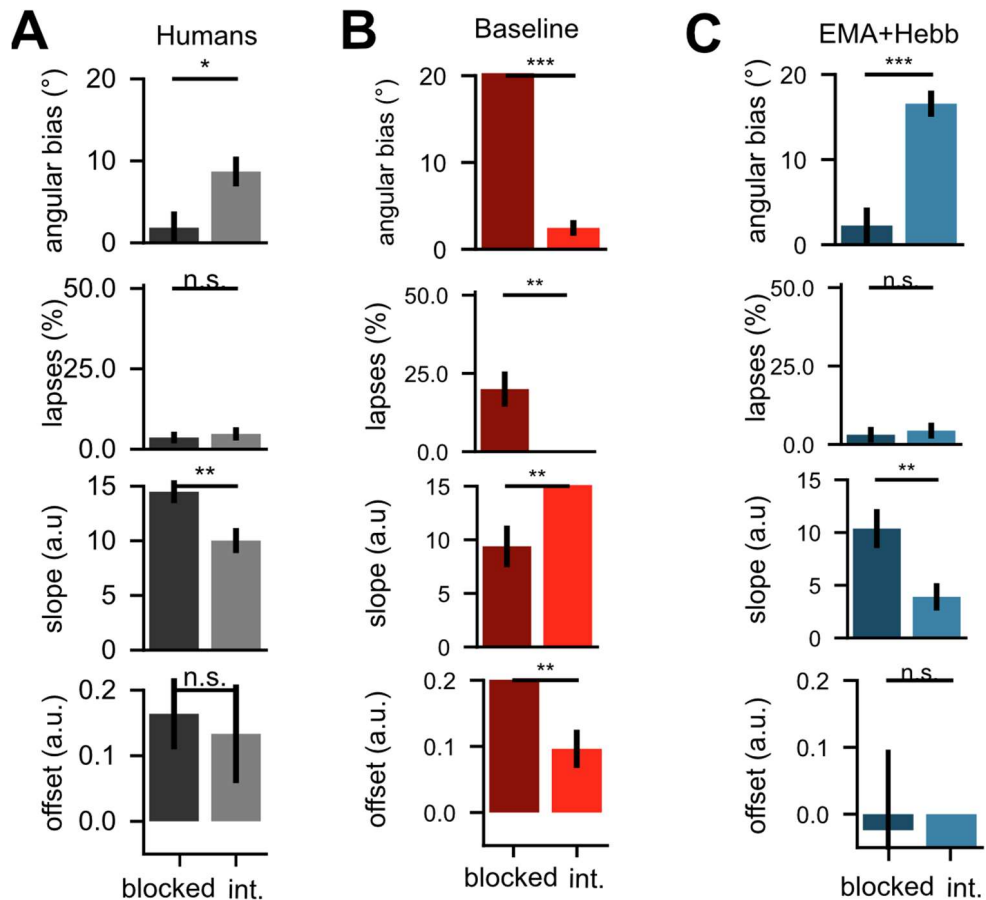

**Figure S7: Fits of Psychometric model to human behaviour and MLPs trained on fractal trees images.** **(A)** Fits of a psychometric model with four parameters to choices made by human participants, decomposing error patterns into (i) angular bias of category boundary (ii) lapse rate (iii) slope and (iv) offset of sigmoidal transducer. Both the angular bias and slope differed significantly between participants trained on blocked and interleaved curricula (bias: T(93) = -2.54, p = 0.0127, slope: T(93) = 2.88, p = 0.0049). **(B)** Same as (A) but fitted to the MLP that was trained on fractal trees images with vanilla SGD. **(C)** Same as (A), but fitted to the outputs of an MLP that was trained on the trees images with sluggishness and additional Hebbian updates. Patterns resembled those observed in human participants (bias: T(98) = -5.47, p < 0.0001, slope: T(98) = 2.86, p < 0.0001).